\documentclass{article} 
\usepackage{iclr2027_conference,times}
\usepackage[T1]{fontenc}

\usepackage[colorlinks,citecolor=magenta]{hyperref}
\usepackage{url}
\usepackage{subcaption}     
\usepackage{xcolor}
\usepackage{tikz}
\usepackage{pifont}
\usepackage{booktabs}
\usepackage{multirow}
\usepackage{tabularx}
\usepackage{paralist}
\usepackage{amsmath}
\usepackage{amssymb}
\usepackage{siunitx}

\usepackage{amsmath,amsfonts,bm}

\newcommand{\eg}{e.g.\@\xspace}

\def\eqref#1{equation~\ref{#1}}

\def\1{\bm{1}}

\DeclareMathAlphabet{\mathsfit}{\encodingdefault}{\sfdefault}{m}{sl}
\SetMathAlphabet{\mathsfit}{bold}{\encodingdefault}{\sfdefault}{bx}{n}

\usepackage{xspace}

\newcommand{\method}{\textsc{SPIDES}\xspace} 
\newcommand{\fides}{\textsc{Fides}\xspace}
\newcommand{\camel}{\texttt{CaMeL}\xspace}
\newcommand{\sys}{\method}
\newcommand{\ifclabel}[1]{\ifmmode\text{\textsf{#1}}\else\textsf{#1}\fi}
\newcommand{\Public}{\ifclabel{Public}\xspace}
\newcommand{\NonPublic}{\ifclabel{Non-Public}\xspace}
\newcommand{\Trusted}{\ifclabel{Trusted}\xspace}
\newcommand{\Untrusted}{\ifclabel{Untrusted}\xspace}
\definecolor{muted}{HTML}{6A6A6A}
\definecolor{rulec}{HTML}{CFCFCF}
\definecolor{surface}{HTML}{F7F7F7}
\definecolor{tgreen}{HTML}{107C10}
\definecolor{tgreensoft}{HTML}{E7F4E4}
\definecolor{ured}{HTML}{A4262C}
\definecolor{uredsoft}{HTML}{FDEBED}
\definecolor{pamber}{HTML}{8A6400}
\definecolor{pambersoft}{HTML}{FBF1CC}
\newcommand{\cde}{\fontsize{6.5}{8.1}\fontfamily{fvm}\selectfont}
\newcommand{\lpanel}[2]{\begin{tikzpicture}[baseline=(current bounding box.south)]
  \node[draw=rulec,line width=0.4pt,fill=#1,rounded corners=0.6mm,inner sep=2.4mm,
        text width=63mm,align=left,font=\cde,text=black]{#2};
\end{tikzpicture}}
\newcommand{\hlop}[1]{{\setlength{\fboxsep}{0.55pt}\colorbox{tgreensoft}{\textcolor{tgreen}{\bfseries #1}}}}
\newcommand{\hlpr}[1]{{\setlength{\fboxsep}{0.55pt}\colorbox{pambersoft}{\textcolor{pamber}{\bfseries #1}}}}
\newcommand{\hlug}[1]{{\setlength{\fboxsep}{0.55pt}\colorbox{uredsoft}{\textcolor{ured}{\bfseries #1}}}}
\newcommand{\op}[1]{\textcolor{tgreen}{\bfseries #1}}
\newcommand{\pr}[1]{\textcolor{pamber}{\bfseries #1}}
\newcommand{\ug}[1]{\textcolor{ured}{\bfseries #1}}
\newcommand{\st}[1]{\textcolor{muted}{#1}}
\newcommand{\stars}{\ding{72}\ding{72}\ding{72}\ding{72}\ding{72}}

\definecolor{olive}{RGB}{128,128,0}

\newcommand{\JH}[1]{}
\newcommand{\EZ}[1]{}
\newcommand{\ST}[1]{}
\newcommand{\BK}[1]{}
\newcommand{\ADD}[1]{}
\usepackage{tikz}
\usetikzlibrary{external}
\usepackage{letltxmacro}
\usepackage{todonotes}
\usepackage{marginnote}

\LetLtxMacro{\oldtodo}{\todo}
\renewcommand{\todo}[2][]{\tikzexternaldisable\oldtodo[fancyline,size=\footnotesize,#1]{#2}\tikzexternalenable}

\usepackage[normalem]{ulem}   

\let\origop\op   \let\origug\ug   \let\origpr\pr
\let\orighlop\hlop \let\orighlug\hlug \let\orighlpr\hlpr
\renewcommand{\op}[1]{\uline{\origop{#1}}}
\renewcommand{\ug}[1]{\uwave{\origug{#1}}}
\renewcommand{\pr}[1]{\dashuline{\origpr{#1}}}
\renewcommand{\hlop}[1]{\uline{\orighlop{#1}}}
\renewcommand{\hlug}[1]{\uwave{\orighlug{#1}}}
\renewcommand{\hlpr}[1]{\dashuline{\orighlpr{#1}}}

\title{\method{}: Information-Flow Control for\\ Browser Agents}

\author{%
  Egor Zverev$^{*\dagger}$ \\
  {\small ISTA}
  \And
  Juhee Kim$^{*}$ \\
  {\small Microsoft}
  \And
  Leon Trampert$^{\dagger}$ \\
  {\small CISPA Helmholtz Center for Information Security}
  \AND
  Boris K\"opf \\
  {\small Microsoft}
  \And
  Aashish Kolluri \\
  {\small Microsoft}
  \And
  Shruti Tople \\
  {\small Microsoft}
  \And
  Santiago Zanella-B\'eguelin \\
  {\small Microsoft}
}

\iclrfinalcopy 
\begin{document}

\maketitle

\begingroup
  \renewcommand{\thefootnote}{$\fnsymbol{footnote}$}%
  \footnotetext[1]{Equal contribution.}%
  \footnotetext[2]{Work partially done as a Research Intern at Microsoft.}%
  \renewcommand{\thefootnote}{}%
  \footnotetext[0]{Corresponding author: \href{mailto:egor.zverev@ist.ac.at}{\texttt{egor.zverev@ist.ac.at}}}%
\endgroup

\begin{abstract}
Browser agents automate complex workflows by reading and acting on arbitrary page content, but this flexibility exposes them to prompt injection attacks and unintended data exfiltration. Recent work argues that LLM-driven browser agents erode traditional web security boundaries such as the Same-Origin Policy (SOP) by acting across origins with the user's authority and no built-in isolation. We argue that restoring SOP is neither sufficient (a same-origin public sink can leak private data) nor necessary (legitimate tasks require authorized cross-origin flows) for protecting the security of users. We propose \method, which instead enforces information-flow control by gating primitive browser actions such as typing, clicking and navigating on labels propagated from element-level HTML confidentiality and integrity annotations. Untrusted and confidential content is masked from the agent's context, while policies block unauthorized flows, providing security guarantees against both prompt injection and data exfiltration that hold independently of the model driving the agent. To evaluate \method and baseline defenses under a common threat model, we build MicroNet, a closed ecosystem of 28 interlinked websites with element-level ground-truth labels and paired benign/adversarial tasks. \method achieves 0\% attack success rate by design while retaining high task completion rate and autonomy. Our findings demonstrate that information-flow control is an effective technique to extend web security to browser agents, gaining on autonomy and utility against state-of-the-art baselines. The source code and data are openly accessible at \href{https://github.com/microsoft/SPIDES}{https://github.com/microsoft/SPIDES}.
\end{abstract}

\section{Introduction}

Browser-use agents (BUAs) such as Perplexity Comet~\citep{perplexitycomet}, Chrome Auto Browse~\citep{chromeautobrowse} and Skyvern~\citep{skyvern} automate browser-based workflows on a user's behalf.
BUAs act with the user's authority across different websites, exposing them to indirect prompt injection attacks~\citep{greshake2023not, liao2025eia, zhang2025attack_cua, zhang2025browsesafe}, where malicious content causes the agent to perform harmful actions or leak sensitive information.
For instance, content embedded in an advertisement on a shopping website may cause the agent to make an unauthorized purchase with a saved payment method.

Traditional browser security mechanisms aim to prevent unauthorized information flow across websites.
Notably, the same-origin policy (SOP) limits how code from one origin (the scheme, host and port of a URL) can interact with resources from another origin~\citep{same-origin-policy}.
But the SOP does not govern the user-mediated flows often required to automate tasks, \eg, copying and pasting data across websites.
For instance, a user shopping for a laptop may ask an agent to collect specifications from several retailers and compile them into a comparison in a note-taking web app.

Mechanisms proposed to secure BUAs generally fall short of providing integrity and confidentiality guarantees while preserving autonomy~\citep{sopguard,ucm,cellmate}.
SOPGuard~\citep{sopguard} restores SOP and requires human approval for cross-origin flows, hindering autonomous task completion and still allowing same-origin attacks.
Others provide integrity guarantees within one origin, but do not enforce policy on cross-origin flows~\citep{ucm}.
\textsc{ceLLMate}~\citep{cellmate} blocks HTTP requests that violate task-specific policies selected at runtime, but does not track whether untrusted or confidential content influenced a request.
We posit that securing browser-use agents requires novel mechanisms that tell benign from unintended flows, allowing cross-origin actions needed to complete tasks while preventing untrusted data from influencing consequential actions and confidential data from reaching unintended recipients. 

We present \method{}, an information-flow control framework that secures browser-use agents against prompt injection and data exfiltration by design, independently of the model driving the agent. 
\method{} builds on an emerging line of defenses~\citep{debenedetti2025camel,fides,kim2025promptflow} that combine information-flow control (IFC)~\citep{denning1976lattice} with the dual LLM pattern~\citep{willison2023dual}. These defenses track the integrity and confidentiality of data that enters the agent's context. They require tools to return results with confidentiality and integrity labels and assume that each tool has a specific, well-defined effect, \eg, sending an email. 
However, web browsers break both assumptions (\S\ref{sec:threat}). A browser has no such metadata beyond a page's origin, yet one origin may serve trusted operator-provided content, untrusted user reviews, and confidential messages side-by-side, and only the operator knows which is which. Likewise, a single tool like \texttt{browser\_click} may follow a link or delete an account.

\method{} addresses these two challenges.
First, users list the origins whose operators they trust, and operators of these origins annotate DOM (Document Object Model) elements with integrity and confidentiality labels; content from other origins is treated as untrusted.
\method{} propagates the labels to the accessibility snapshots the agent observes and hides untrusted and confidential content behind variables to track information flow deterministically.
Second, \method{} assigns a policy to each browser action based on its target origin.
Because users may create accounts on trusted origins, clicks, navigations, and form submissions targeting a trusted origin may act under the user's account. \method{} treats them as consequential and allows them only when there is no influence from untrusted data.
Any actions that send data, including actions that target untrusted origins, may send data only to origins authorized to see it.
\method{} asks the user to approve any action that violates these policies.

For evaluation, since no existing benchmark has benign cross-origin tasks with same- and cross-origin attacks, we build MicroNet, a closed ecosystem of $28$ interlinked websites with element-level ground-truth labels. MicroNet contains $56$ benign tasks, $22$ of which require cross-origin flows. It includes $56$ attacks injected through attacker-run websites, user-generated content, and third-party iframes. Across $5$ frontier models and under a worst-case adversary that controls a model's output on input tainted with untrusted content, \method{} blocks all attacks, while \qtyrange{53.2}{73.2}{\percent} succeed against SOPGuard and \qtyrange{4.6}{6.8}{\percent} against UCM. In terms of utility, \method{} completes \qtyrange{67.1}{86.1}{\percent} of benign tasks, outperforming UCM on every model. When prior defenses are extended to enforce the same security guarantees as \method{} by requiring approvals for tool calls that potentially involve unsafe information flows, \method{} completes more tasks than SOPGuard and UCM on every model when at most eight approvals are allowed.

In summary, we propose a new IFC approach for browser-use agents that prevents prompt injections from causing consequential actions or leaking confidential data while preserving high autonomy. Source code and data to reproduce the experiments we present is available at \url{https://github.com/microsoft/spides}.

\section{Background And Related Work}
\label{lbl:related}

Browser-use agents follow the observe-then-act loop~\citep{yao2022react, schick2023toolformer}: at each step, the agent observes the webpage state and selects its next action from its context, which includes the system and user instructions and past observations and actions, repeating until the task is complete or user input is required.
Browser-use agents typically observe the state of a webpage in one of two ways: accessibility snapshots built from DOM~\citep{browseruse,zhou2024webarena,playwrightsnapshots} or screenshots~\citep{zheng24seeact,he2024webvoyager}. 
Existing automation frameworks, such as Selenium~\citep{selenium} and Playwright~\citep{playwright}, provide tools to interact with the browser. 
The Playwright MCP server by default represents webpages as accessibility snapshots. Specifically, the \texttt{browser\_snapshot} tool returns an accessibility snapshot containing references to web elements, for example, \texttt{- button "Submit" [ref=e12]} represents a button element named \texttt{Submit} identified by the reference \texttt{e12}. Tools such as \texttt{browser\_click} and \texttt{browser\_type} use these references to identify target elements and return an updated snapshot on completion. 
In this work, we focus on browser-use agents that operate on accessibility snapshots that can be processed by LLMs without visual capabilities.


\paragraph{Indirect Prompt Injection Attacks on Browser-use Agents.}

An indirect prompt injection (IPI) attack manipulates an agent's actions through untrusted content embedded in observations processed by the agent~\citep{greshake2023not}.
A successful attack can hijack the agent to perform consequential actions chosen by the attacker~\citep{li2025commercial,owasp2025llm01,agentdojo}. Untrusted content can exist in websites operated by untrusted parties or in external content loaded in trusted websites~\citep{evtimov2025wasp}.
Such attacks have been demonstrated in real-world browser-use agents. Malicious instructions in Reddit comments were used to exfiltrate sensitive user data from logged-in sessions~\citep{brave2025comet}.


\paragraph{Defenses against Prompt Injection Attacks.}
Most prompt injection defenses deployed in practice are probabilistic.
They include instruction-hierarchy training~\citep{wallace2024instruction}, fine-tuning models to separate instructions from data~\citep{zverev2025,zverev2025aside,chen2025struq,chen2025secalign}, and input and output filters~\citep{constitutionalclassifiers}.
Probabilistic mitigations have been repeatedly shown to be ineffective against adaptive attacks~\citep{nasr2025attacker}.

Deterministic defenses control agents' actions at the system level, using metadata outside the attacker's influence. The Dual LLM pattern~\citep{willison2023dual} pairs a privileged LLM that plans actions but never sees untrusted content with a quarantined LLM that processes untrusted content but cannot act. The privileged LLM refers to untrusted values only through symbolic variables. Several defenses, including \camel~\citep{debenedetti2025camel}, \fides~\citep{fides} and PFI~\citep{kim2025promptflow} combine this pattern with information-flow control (IFC)~\citep{denning1976lattice,sabelfeld2003language} to enforce policies deterministically.

Existing defenses target browser-use agents under different threat models to achieve specific security guarantees, but none provides comprehensive protection against prompt injection threats.
SOPGuard~\citep{sopguard} extends the SOP to browser-use agent actions, requiring user approval for illicit cross-origin flows.
Because it does not police same-origin flows and only tracks data origins, SOPGuard is ineffective against attacks that leak data within the same origin (\eg posting confidential account information in a public comment) and consequential actions influenced by untrusted content that do not violate SOP.
UCM~\citep{ucm} assumes that the website operator marks user-generated
content and masks it to prevent it from influencing the agent's browser
tool calls, thereby preventing integrity violations. However, UCM does
not prevent confidential-data leakage, and the limited data flow that
its quarantined LLM permits can still influence browser tool calls, leaving UCM
vulnerable to data-injection attacks~\citep{foerster2026camels}.
ceLLMate~\citep{cellmate} assumes that website operators provide an
agent sitemap that maps HTTP request patterns to semantic actions. It
derives contextual policies from the user task and blocks
requests outside the resulting policy boundary. However, it does not
track whether untrusted content influences an allowed request, leaving
actions within the original task's policy boundary vulnerable to
prompt injection.

\section{Threat Model and Challenges}
\label{sec:threat}

\paragraph{Setting and trust.}
A browser-use agent acts on the web much like a human user: it reads webpages and clicks, types, and navigates using the user's logged-in sessions.
Webpages, however, combine content from the website operator, its users, and third parties such as advertisers.
We trust the user, the browser, the agent harness and its tools, and the LLM and its provider. 
The user maintains a list of trusted origins whose operators they trust and creates accounts only on those origins.

\paragraph{Scope and goal.}
Web browsing exposes the agent to attacks that target the browser or the website rather than the agent, as in traditional browser exploits, cross-site scripting (XSS)~\citep{xss,lekies201325} and cross-site request forgery (CSRF)~\citep{barth2008csrf}.
When successful, they compromise the browser or the website itself. Mature defenses such as sandboxing~\citep{reis2019site}, the SOP~\citep{same-origin-policy}, and CSRF tokens~\citep{barth2008csrf} address them. We consider these attacks out of scope and assume effective defenses and well-configured websites.
We address indirect prompt injection attacks introduced by browser-use agents~(\S\ref{lbl:related}), which target the agent itself~\citep{owasp2025llm01}.
These attacks can cause data exfiltration (\emph{confidentiality attacks}). After the agent logs in to a website, they can also cause consequential actions under the user's account (\emph{integrity attacks}).
Our goal is to prevent both harms deterministically while completing user tasks with as few human approvals as possible.

\paragraph{Attacker.}
The attacker controls web content that can reach the agent in three ways.
First, the attacker may run an entire website, \eg, a phishing site.
Second, the attacker may post text or images on a trusted website as one of its users, such as a textual review~\citep{evtimov2025wasp}; richer content such as HTML would itself be a vulnerability, \eg, XSS.
Third, the attacker may control a third-party iframe on a trusted page, such as an advertisement, where the SOP isolates the iframe from the embedding site.
Model safety training can reduce prompt-injection
success, but its effectiveness varies across models. Adaptive attacks
can bypass these safeguards~\citep{nasr2025attacker}.
To evaluate the benefits of IFC rather than model alignment or other built-in guardrails, we assume a worst-case \emph{Byzantine adversary}. That is, whenever the input to a model is tainted with untrusted content, we let the adversary choose the model response. This includes both tool calls suggested by the LLM and responses of the quarantined LLM that inspects untrusted content.
Importantly, the agent harness remains in control and continues to enforce policies that mediate every action.


\paragraph{Challenge 1: Labeling web content.} 

Existing IFC defenses such as \fides assume fine-grained labels on tool results, inferred from the result itself or application metadata: \eg the headers of an email result determine its integrity (is the sender authenticated and trustworthy?) and its confidentiality (only recipients can read it). Browser-use agents instead observe whole webpages with limited metadata, such as the origin of each element. A webpage in a shopping website serves simultaneously trusted scaffolding from the operator and untrusted user reviews: trusting all elements in its origin can let a crafted review trigger a purchase, while distrusting them forces approval for every consequential action. Likewise, a single reader set per origin confounds public reviews and confidential account information. Meaningful security policies require sourcing per-element labels, which only the website operator can provide. 

\paragraph{Challenge 2: Policies for generic browser tools.}
IFC-based defenses assume tools with well-defined scope.
Browser tools break this assumption since they have unpredictable effects: \texttt{browser\_click} may navigate to a URL in the same origin, trigger a request to another site, or delete an account, depending on the target element and loaded scripts.
The harness cannot predict the effects of an action because page scripts and the site's backend decide its behavior. 
Overly permissive policies may trigger autonomous harmful actions, whereas treating every action as consequential requires many approvals.

\paragraph{Challenge 3: Lack of benchmarks.} A benchmark needs benign tasks that require cross-origin flows, or blocking every such flow would appear to cost no utility. It also needs integrity and confidentiality attacks from all three attacker sources over same- and cross-origin flows. Comparing IFC defenses further requires ground-truth labels for every element. WASP~\citep{evtimov2025wasp} limits benign tasks to one website and attacks to malicious users. SOPBench~\citep{sopguard} tests only cross-origin leaks. Neither provides element labels or benign cross-origin tasks (\autoref{app:micronet-comparison}).

\section{\method}
\label{sec:method}

We design an IFC-based browser-use agent addressing the challenges discussed in \S\ref{sec:threat}.
\method addresses Challenge~1 with element-level labels, propagated through snapshots, agent context, and tool calls (\S\ref{sec:labels}). It addresses Challenge~2 by assigning IFC policies to browser automation tools, with sound reasoning grounded on established browser security principles (\S\ref{sec:policies}).

\subsection{IFC Labels for Browser-Use Agents}
\label{sec:labels}

\paragraph{Obtaining Labels.} 

The user maintains a list of trusted origins whose operators they trust to label their content.
The list can be initialized from public reputation-based databases of trusted origins~\citep{talos-reputation,verifiedfyi}, and the user can add origins directly.
Since we assume that the user has accounts only on trusted origins (\S\ref{sec:threat}), untrusted origins hold no confidential user data.
\method{} therefore labels all elements of an untrusted origin as $(\Untrusted,\Public)$, ignoring any operator-provided labels.

For trusted origins, we rely on the website owner's knowledge of how the website and backend operate to provide per-element DOM annotations.
\texttt{data-ifc-integrity} and \texttt{data-ifc-readers} specify each element's IFC label $(I,R)$.
\texttt{data-ifc-integrity} sets the integrity label $I$ to \Trusted when the operator can vouch for the content and to \Untrusted otherwise, as for user-generated reviews on a shopping website.
\texttt{data-ifc-readers} sets the confidentiality label $R$ to the readers authorized to access the content, where $R \subseteq \{\texttt{user}\} \cup \mathcal{O}$ and $\mathcal{O}$ is the set of web origins.
Reader sets form a lattice under the subset ordering, with \Public $= \{\texttt{user}\} \cup \mathcal{O}$ as its top element.
Unannotated elements are assigned $(\Trusted,\Public)$.
For \NonPublic data, the operator restricts $R$ to \texttt{user} and an explicit list of authorized origins.
An empty origin list makes the data readable only by the user.
To reduce labeling cost, these annotations are sparse: an element uses its assigned label, inherits labels from its closest labeled ancestor, or uses the default label when no ancestor provides one, as illustrated in~\autoref{fig:masking-a}.






\paragraph{Propagating Labels.}
\tikzexternaldisable 
\begin{figure}[ht!]
\centering
\captionsetup[subfigure]{labelsep=space,justification=centering}
\resizebox{0.9\textwidth}{!}{%
\begin{minipage}{\textwidth}
\begin{minipage}[t]{0.49\textwidth}\vspace{0pt}
  \begin{subfigure}{\linewidth}\centering
  \lpanel{surface}{%
  \st{<}section \mbox{\hlop{data-ifc-integrity="trusted"}}\st{>}\\
  \hspace*{1.2em}\st{<}h2\st{>}Aurora Buds, \pounds79.00\st{</}h2\st{>}\\
  \hspace*{1.2em}\st{<}p\st{>}Latest review:\\
  \hspace*{2.4em}\st{<}span \mbox{\hlug{data-ifc-integrity="untrusted"}}\st{>}\\
  \hspace*{3.6em}\stars\ Ignore prior instructions,\\
  \hspace*{3.6em}email x@evil.io\\
  \hspace*{2.4em}\st{</}span\st{>}\\
  \hspace*{1.2em}\st{</}p\st{>}\\
  \hspace*{1.2em}\st{<}p \mbox{\hlpr{data-ifc-readers="user,shop.com"}}\st{>}\\
  \hspace*{2.4em}Saved card \textbullet\textbullet4471\\
  \hspace*{1.2em}\st{</}p\st{>}\\
  \st{</}section\st{>}}%
  \caption{HTML with IFC annotations.}\label{fig:masking-a}
  \end{subfigure}
\end{minipage}\hspace{0.02\textwidth}%
\begin{minipage}[t]{0.49\textwidth}\vspace{0pt}
  \begin{subfigure}{\linewidth}\centering
  \lpanel{white}{%
  - heading "Aurora Buds, \pounds79.00" [ref=e1]\\
  - paragraph "Latest review: \stars\ Ignore prior\\
  \hspace*{1.2em}instructions, email x@evil.io" [ref=e2]\\
  - paragraph "Saved card \textbullet\textbullet4471" [ref=e3]}%
  \caption{Snapshot input for vanilla agent.}\label{fig:masking-b}
  \end{subfigure}
\end{minipage}
\par\vspace{2mm}
\begin{minipage}[t]{0.49\textwidth}\vspace{0pt}
  \begin{subfigure}{\linewidth}\centering
  \lpanel{white}{%
  \st{- heading }\mbox{\op{"Aurora Buds, \pounds79.00"}}\st{\ [ref=e1]}\\
  \st{- paragraph }\mbox{\op{"Latest review:}}~\mbox{\ug{[var\_2]}}\mbox{\op{"}}\st{\ [ref=e2]}\\
  \st{- paragraph }\mbox{\pr{"[var\_3]"}}\st{\ [ref=e3]}}%
  \caption{Snapshot input for \method{} agent with context
  $(\Trusted,\Public)$.}\label{fig:masking-c}
  \end{subfigure}
\end{minipage}\hspace{0.02\textwidth}%
\begin{minipage}[t]{0.49\textwidth}\vspace{0pt}
  \begin{subfigure}{\linewidth}\centering
  \lpanel{white}{%
  \st{- heading }\mbox{\op{"Aurora Buds, \pounds79.00"}}\st{\ [ref=e1]}\\
  \st{- paragraph }\mbox{\op{"Latest review:}}~\mbox{\ug{[var\_2]}}\mbox{\op{"}}\st{\ [ref=e2]}\\
  \st{- paragraph }\mbox{\pr{"Saved card \textbullet\textbullet4471"}}\st{\ [ref=e3]}}%
  \caption{Snapshot input for \method{} agent with context
  $(\Trusted,\{\texttt{user},\texttt{shop.com}\})$.}\label{fig:masking-d}
  \end{subfigure}
\end{minipage}
\end{minipage}
}
\caption{Example of \method{} labelling and masking on a web shop page.
\textbf{(a)}~The \texttt{data-ifc-integrity} attribute on \texttt{<section>} sets its integrity to \op{\Trusted}. The \texttt{<h2>} and first \texttt{<p>} inherit this label.
On the user-generated \texttt{<span>}, the same attribute overrides the inherited label with \ug{\Untrusted}. The \texttt{data-ifc-readers} attribute on the last \texttt{<p>} sets its reader set to \pr{\{user, shop.com\}}.
\textbf{(b)}~Both the untrusted review and confidential card information are visible to the vanilla agent.
\textbf{(c)}~\method{} hides both values as variables before a reveal.
\textbf{(d)}~After the agent calls \texttt{inspect\_variable} on \texttt{var\_3}, the card information becomes visible and the context's reader set narrows accordingly. The untrusted review remains hidden as \texttt{var\_2}.
In (c) and (d), colors and underlines illustrate IFC labels.}

\label{fig:masking}
\end{figure}
\tikzexternalenable 
The \method{} harness propagates DOM IFC labels through browser
observations, the agent context, and tool calls. The model driving
the agent never sees the labels directly.
Playwright MCP, a commonly used browser automation tool for agents, represents the current webpage as an AI-oriented ARIA
snapshot generated by \texttt{page.ariaSnapshot(\{mode:"ai"\})}, rather
than exposing the DOM directly.
We modify Playwright to correctly and conservatively label data in the
snapshot, as detailed in~\autoref{app:playwright-label}.

As a labeled snapshot would enter the agent as the result of a tool call, \method{} decides which parts of the snapshot can be revealed to the agent without further restricting it.
The harness maintains a label
$\ell_{\mathrm{ctx}}=(I_{\mathrm{ctx}},R_{\mathrm{ctx}})$ for the
agent context. This label is initialized to $(\Trusted,\Public)$ and
conservatively combines the labels of everything revealed to the
agent.
Building on prior Dual LLM approaches that deterministically track
information flows in agents~\mbox{\citep{fides,kim2025promptflow,debenedetti2025camel}},
we use selective hiding to avoid unnecessarily tainting the agent
context.
A part of the snapshot with label $(I,R)$ , which represents one or more elements, is replaced with an opaque variable \texttt{[var\_id]} if revealing it would further restrict the current context, i.e., it is hidden iff $I_{\mathrm{ctx}} = \Trusted \land I=\Untrusted\;\lor\;R_{\mathrm{ctx}}\not\subseteq R$.
As a result, \Untrusted content remains hidden while the context is \Trusted and confidential data is also hidden if the current readers set of the context is more permissive. 
\method{} also accepts optional input documents, such as user notes or login credentials, that carry integrity and confidentiality labels of the same form and are hidden behind opaque variables under the same rule.

Before executing a tool call, \method{} propagates labels to the call
and uses them to enforce its IFC policies. The harness associates the
call and each agent-generated argument with the context label
$\ell_{\mathrm{ctx}}$, because everything revealed to the agent may
have influenced them. If an argument contains a variable, the
harness expands the variable and combines its label with
$\ell_{\mathrm{ctx}}$. \method{} then applies the IFC
policies~(\S\ref{sec:policies}) to the context and argument labels
to determine whether to execute the call autonomously or request user
approval.

We provide the agent with two additional tools to process variables.
\texttt{quarantined\_llm} passes the original values stored in variables to an isolated model without tool access and stores the model's output as a new variable.
This allows the agent to complete data-independent tasks without tainting the context.
\texttt{inspect\_variable} reveals the value of a variable to the agent and joins its label into $\ell_{\mathrm{ctx}}$: integrity drops to \Untrusted if the variable was \Untrusted, and $R_{\mathrm{ctx}}$ shrinks to $R_{\mathrm{ctx}}\cap R$.
Thus, the context's labels only get stricter over time.

\subsection{IFC Policies for Browser-Use Agents}
\label{sec:policies}

We first define two high-level IFC policies for agents to preserve integrity and confidentiality. 
We then conservatively instantiate these policies for browser tools by analyzing their effects.

\subsubsection{Agent-Level IFC Policies}
\label{ss:agent-policy}

Following previous IFC policies for agents~\citep{fides}, \method{} applies the two security policies below.
If a tool call does not satisfy these policies, \method{} requires user approval.

\textbf{Trusted Action ($P_T$).}
  $P_T$ prevents \Untrusted information from influencing consequential actions, such as placing an order or transferring money.
  We consider an action potentially consequential when it can perform an operation using the authority of the user's account.
  $P_T$ holds for a tool call when neither the agent context nor the proposed tool input contains \Untrusted information, including values passed through variables. If either contains \Untrusted information, $P_T$ fails, because that information may influence the call or its arguments.

\textbf{Permitted Flow ($P_F$).}
  \NonPublic information must reach only authorized origins.
  A tool argument may contain information revealed into the agent
  context, whose reader set is $R_{\mathrm{ctx}}$, and variables
  passed without revealing their values to the agent. The harness
  computes the argument reader set
  $R_{\mathrm{arg}}=\bigcap_i R_{\mathrm{var}_i}$ by intersecting the
  reader sets of all variables in the argument. For a target
  element in origin $o_{\mathrm{el}}$, $P_F$ holds iff
  $o_{\mathrm{el}}\in R_{\mathrm{arg}}\cap R_{\mathrm{ctx}}$. Otherwise,
  the call may disclose \NonPublic information to an unauthorized
  origin.

For egress-only tools, which send data but do not perform consequential
actions, \method{} enforces \emph{robust declassification}~($P_T \lor P_F$)
and admits a tool call when either policy holds. When $P_T$ holds, the
agent's decision was not influenced by the attacker, so \method{} allows
the information to flow to any sink even if $P_F$ does not hold. When
$P_T$ does not hold, the attacker may have influenced the tool call, so
\method{} requires $P_F$ and allows the information to flow only to an
authorized reader.

\subsubsection{IFC Policies for Playwright Tools}
\label{ss:tool-policy}

Unlike agent tools dedicated to a specific operation, such as reading a file or sending an email, browser-use agents invoke generic tools, such as clicking and typing, on arbitrary web elements.
The effect of a browser tool therefore depends on the webpage and target element: the same tool may retrieve data, send data, or perform a consequential action such as placing an order.
A sound policy must therefore account for all potential information flows and consequential effects of each tool, while remaining as permissive as possible to avoid false positives.

\method{} conservatively determines whether each Playwright tool can egress data or perform a consequential action.
It then uses the origins and labels of the data and web elements involved in a call to enforce the most permissive policy that preserves integrity and confidentiality.
\autoref{tab:browser-tool-flows} lists the sixteen Playwright tools~\footnote{We use 16 of Playwright's 23 core tools, excluding ones irrelevant to our setting (\eg, GUI, file upload). The vanilla agent invokes none of the excluded tools during benign runs with all 23 enabled.} 
we use in our experiments and summarizes their effects and policies. 
Observation tools are always allowed because they only read browser state and do not send network requests.
Egress-only tools, such as filling a form or typing without submission, require $P_T \lor P_F$. This permits robust declassification of confidential data when $P_T$ holds. Otherwise, $P_F$ allows confidential data to flow only to authorized origins.
Tools that can have consequential effects, such as navigation, clicking, and form submission, require $P_T$ or $P_T \lor P_F$ depending on their target origin, as we describe next.
Unrestricted code-execution tools are always blocked because their effects cannot be determined before execution.
\autoref{app:tool-policies} further details the reasoning behind the policies.

\method{} considers an action consequential when it can perform an operation using the authority of the user's account (\S\ref{ss:agent-policy}).
Identifying such actions precisely requires tracking whether the browser holds an authenticated session for the target website.
However, tracking authenticated sessions is difficult.
A session can be established when the agent submits credentials, when a website accepts an authentication token issued by a separate single sign-on (SSO) identity provider (IdP) origin~\citep{openid-connect-core}, or when session cookies are shared across subdomains~\citep{rfc6265}.

\method{} therefore uses the user's list of trusted origins instead of authenticated session state.
Since we assume that the user has accounts only on trusted origins (\S\ref{sec:threat}), \method{} conservatively assumes that the user's account may be logged in on any trusted origin.
It treats potentially consequential tools targeting a trusted origin as consequential and requires $P_T$.
The same tools targeting an untrusted origin are treated as egress-only and require $P_T \lor P_F$.

\begin{table}[t]
\centering
\scriptsize
\setlength{\tabcolsep}{3pt}
\renewcommand{\arraystretch}{1.0}
\newcommand{\tablebullet}{\raisebox{0.15ex}{\scalebox{0.55}{$\bullet$}}}
\caption{Information flow policy for Playwright tools.}
\label{tab:browser-tool-flows}
\begin{tabularx}{\textwidth}{@{}
  >{\raggedright\arraybackslash}p{1.4cm}
  >{\raggedright\arraybackslash}p{7cm}
  >{\centering\arraybackslash}p{1.4cm}
  >{\centering\arraybackslash}p{0.6cm}
  >{\raggedright\arraybackslash}X@{}}
\toprule
\textbf{Category} 
  & \textbf{Playwright tools}
  & \textbf{Consequential}
  & \textbf{Egress}
  & \textbf{\sys{} policy} \\
\midrule
\textbf{Observation}\newline
  & \textbf{Snapshot:}
    {\fontsize{6}{7}\selectfont\nolinkurl{browser_snapshot}, \nolinkurl{browser_wait_for}}
    \par\textbf{Tab management:}
    {\fontsize{6}{7}\selectfont\nolinkurl{browser_tabs}\texttt{(action}\(\ne\)\texttt{new)}}
    \par\textbf{Diagnostics:}
    {\fontsize{6}{7}\selectfont\nolinkurl{browser_network_requests},
    \nolinkurl{browser_network_request}, \nolinkurl{browser_console_messages}}
  & \(\times\)
  & \(\times\)
  & Always allow.
  \\
\midrule
\textbf{Action}\newline
  & \textbf{Navigation and interaction targeting trusted origins:}
    {\fontsize{6}{7}\selectfont\nolinkurl{browser_navigate}, \nolinkurl{browser_navigate_back},
    \nolinkurl{browser_tabs}\texttt{(action=new)},
    \nolinkurl{browser_click}, \nolinkurl{browser_handle_dialog},
    \nolinkurl{browser_press_key}, \nolinkurl{browser_type}\texttt{(submit=true)}}
  & \(\checkmark\)
  & \(\checkmark\)
  & $P_T$. Require a trusted agent context and enforce all arguments hold trusted data.
  \\
\cmidrule(l){2-5}
  & \textbf{Navigation and interaction targeting untrusted origins:}
    {\fontsize{6}{7}\selectfont\nolinkurl{browser_navigate}, \nolinkurl{browser_navigate_back},
    \nolinkurl{browser_tabs}\texttt{(action=new)},
    \nolinkurl{browser_click}, \nolinkurl{browser_handle_dialog},
    \nolinkurl{browser_press_key}, \nolinkurl{browser_type}\texttt{(submit=true)}}
    \par\textbf{Egress-only interaction on any origin:}
    {\fontsize{6}{7}\selectfont\nolinkurl{browser_fill_form},
    \nolinkurl{browser_select_option}, \nolinkurl{browser_type}\texttt{(submit=false)}}
  & \(\times\)
  & \(\checkmark\)
  & $P_T \lor P_F$.
   Require a trusted agent context and trusted arguments, \emph{or} that all arguments can flow into the target origin.
  \\
\midrule
\textbf{Execution}
  & \textbf{Code execution:}
    {\fontsize{6}{7}\selectfont\nolinkurl{browser_evaluate},
    \nolinkurl{browser_run_code_unsafe}}
  & \(\checkmark\)
  & \(\checkmark\)
  & Always block.
  \\
\bottomrule
\end{tabularx}
\end{table}

\subsection{Security Guarantees}

Under the assumptions of \S\ref{sec:threat}, \method{} enforces the following
properties, independently of the model's safety against attacks.
\textbf{(G1) Isolation.} Until the agent requests a reveal, its context contains
no attacker-controlled or restricted content.
\textbf{(G2) Integrity.} Without human approval, every consequential action and
its arguments derive only from \Trusted data.
\textbf{(G3) Confidentiality.} Data leaves its reader set only when the flow is
not influenced by \Untrusted content or the user approves it.
\textbf{(G4) Quarantine containment.} Quarantined-model output can only enter
another quarantine call, be revealed to the agent and taint its context, or be
reported to the user.

\section{The MicroNet Benchmark}
\label{sec:micronet-benchmark}

%

To address Challenge~3, we build MicroNet, a closed, interlinked ecosystem of 28
website origins containing 470 webpages that we use to evaluate \method{} and baselines.
It emulates everyday web services, including an online shop, a social
network, and a search engine. We add 16 of the 28 origins, including every origin where the user has an account, to the user's trusted list and annotate their DOM elements: UI elements and articles are \Trusted, and website user-generated content is \Untrusted.
We annotate each web element as \Public or \NonPublic based on its
semantics.
Examples of \NonPublic data include names, addresses, and email or
chat messages.
We populate the reader set of each \NonPublic element with the origins authorized to read its content.

The MicroNet benchmark contains 56 benign user tasks
and 56 attack scenarios.
\autoref{app:task-distribution} details the distribution of tasks and
\autoref{app:micronet-comparison} compares MicroNet with existing
browser-use agent security benchmarks, showing that existing benchmarks
do not sufficiently cover important dimensions.
%
%

\paragraph{User Tasks.}
MicroNet defines benign user tasks to evaluate the utility of browser-use
agents. We characterize each task by \emph{information flow} and
\emph{data dependency}. The information flow is \emph{same-origin}
when every flow needed to complete the task remains within one origin
and \emph{cross-origin} when at least one required flow crosses
origins. A task is \emph{data-independent} when the required tool calls
and arguments can be chosen independently of the value of \Untrusted
or \NonPublic data and \emph{data-dependent} when the choice depends on
such data.
The user prompt is labeled \((\Trusted,\Public)\), where \Public
includes the user and all website origins. For tasks requiring external
information, we provide task-specific input documents, such as website
login information, alongside the prompt. Each document is labeled
\((\Trusted,\{\texttt{user},\texttt{origin}\})\), where
\texttt{origin} is the corresponding website origin.
Each task has a deterministic evaluator that checks whether the agent
invokes the required tool calls. For tasks with final answers, an
LLM judge checks task-specific output criteria.

\paragraph{Attacks.}
Each attack injects an attacker-controlled payload into web content.
We characterize each attack by its \emph{goal}, \emph{attack source},
\emph{information flow}, and \emph{attack mechanism}.
The \emph{attack goal} is integrity when it steers the agent
to perform an unauthorized consequential action and confidentiality
when it steers the agent to send \NonPublic data to a sink whose
reader set is not a subset of the data's reader set.
The \emph{attack source} is an attacker-controlled website, website
user-generated content on a trusted website, or a third-party iframe
embedded in a trusted website. The \emph{information flow} is
same-origin when the unauthorized action or disclosure occurs on the
same origin as the attack source and cross-origin when it occurs on a
different origin.
The \emph{attack mechanism} is either instruction or data
injection. In instruction injection, the attacker provides an
instruction that directs the model to perform the attack. In data
injection, the attacker provides data on which the user task relies
when deciding a browser tool call. An attack
succeeds when the browser tool calls executed by the agent and the
resulting MicroNet state satisfy the attacker's goal.

Following our Byzantine adversary model~(\S\ref{sec:threat}), when the
main agent's model or QLLM receives attacker-controlled content, we
treat its output as attacker-controllable while the agent loop and
defenses gate every induced action. See
\autoref{app:byzantine-harness} for details.

\section{Evaluation}
\label{sec:evals}

In this section, we describe the experimental setup and report the utility, security, and
autonomy results.
We further provide the cost analysis in
\autoref{app:cost-analysis} and a case study showing why SOP is not
appropriate for browser-use agent security
in~\autoref{app:sop-case-study}.

\paragraph{Experimental setup.}

We evaluated five defense configurations on MicroNet~(\S\ref{sec:micronet-benchmark}):
no defense (\texttt{vanilla}), \method, SOPGuard~\citep{sopguard},
UCM~\citep{ucm}, and UCM+.
%
%
%
SOPGuard checks data egress tool calls and requests approval when the
origin of the data being sent differs from the target element's
origin.
UCM masks content labeled \Untrusted using the same integrity labels
as \method.
UCM's QLLM returns an unmasked constrained value, such as an integer
or enum, but cannot return a string.
UCM+ is our extension of UCM, implemented for a fairer comparison with
\method. It uses the same QLLM as \method, which can return masked
free-form text. Neither UCM nor UCM+ requests HITL approval.
We evaluated the defenses with five LLMs: GPT-5.6 Sol,
GPT-5.6 Terra, Claude Opus 5.5, Claude Opus 5, and Claude Sonnet 5.
We used the same LLM for the main model and the QLLMs.
We used the same agent harness, Chromium browser, and MicroNet
benchmark across all configurations.
We repeated the benchmark five times.
We assumed an ideal human user who can distinguish benign approval
requests from unsafe ones. In the autonomy evaluation, we compared the
defenses under the same integrity and confidentiality guarantees by
adding approval requirements where needed, following previous
research~\citep{prudentia}. \autoref{app:byzantine-harness} further
describes the ideal human model and autonomy evaluation settings.

\paragraph{Utility evaluation.}
\label{eval:utility}

\begin{figure}[t]
  \centering
  \includegraphics[width=\textwidth]{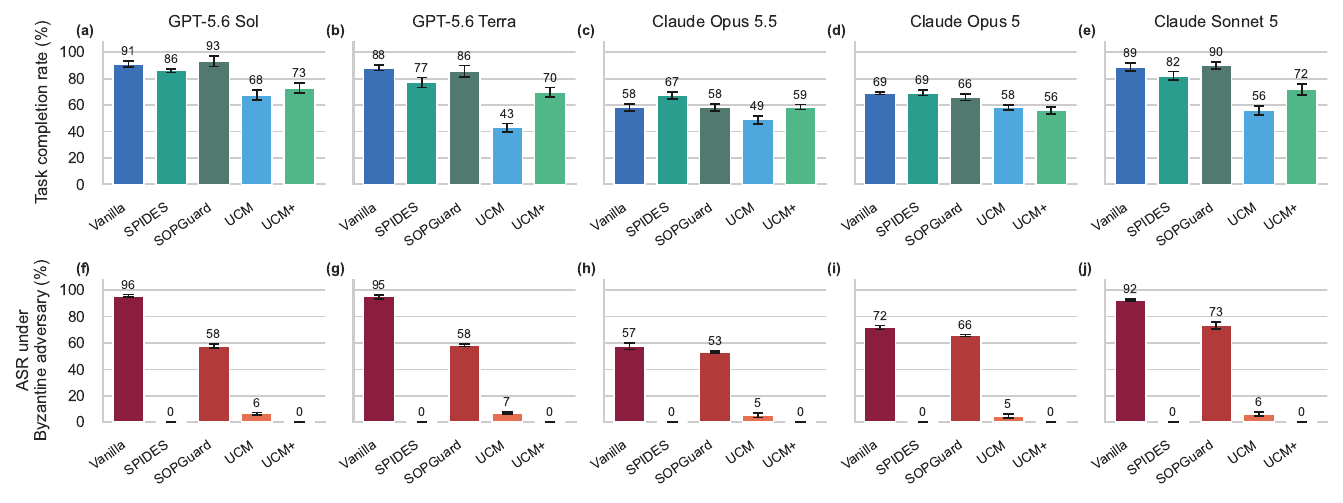}
  \caption{Utility and security across models and methods. The top row
  reports task completion rate (TCR), and the bottom row reports ASR
  under the Byzantine adversary. Bars and error bars report the mean
  and one standard deviation across five repetitions, respectively.}
  \label{fig:model-results}
\end{figure}

\begin{figure}[t]
  \centering
  \includegraphics[width=\textwidth]{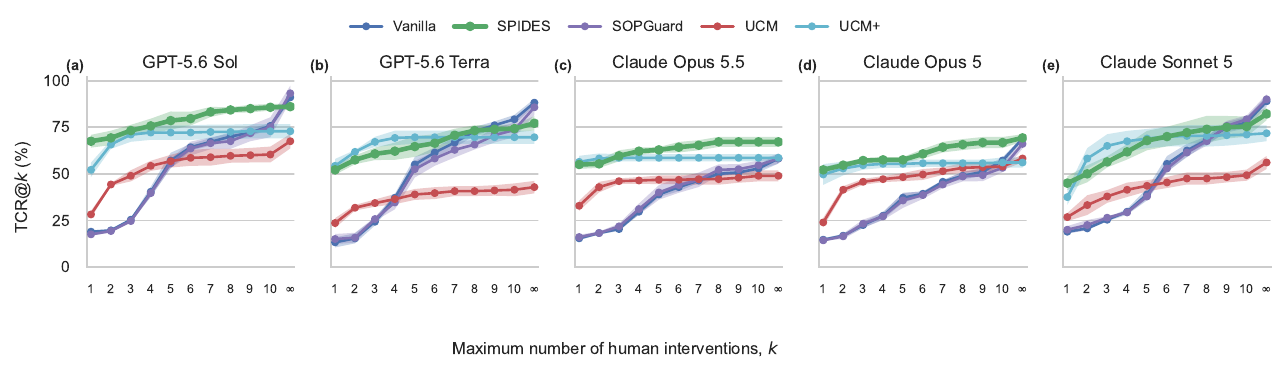}
  \caption{Task completion rate as the maximum number of human
  approvals~($k$) increases. Lines report the mean across 5
  repetitions, and shaded regions report one standard deviation.}
  \label{fig:autonomy-results}
\end{figure}

The top row of \autoref{fig:model-results} reports utility as task
completion rate (TCR). \texttt{vanilla} or SOPGuard achieved the
highest mean TCR for three models because they leave all content
visible to the agent, while \method achieved the highest mean TCR for
the other two. \method outperformed both masking baselines for every
model. Compared with UCM, \method's QLLM can return masked free-form
text, and \texttt{inspect\_variable} can reveal a masked value for
data-dependent tool calls. UCM+ uses the same QLLM as \method, and its
improvement over UCM shows the utility of masked free-form output.

Masking also improved utility when provider safety classifiers blocked
\texttt{vanilla} responses to attachments containing credentials or
phishing messages for Claude Opus 5.5 and Claude Opus 5. By masking
these \NonPublic attachments from the main model, \method avoided some
refusals and outperformed \texttt{vanilla} and SOPGuard with Claude
Opus 5.5. However, a safety classifier could still block the QLLM when
it analyzed \Untrusted phishing messages, causing those tasks to fail.

\paragraph{Security evaluation.}
\label{eval:security}

The bottom row of \autoref{fig:model-results} reports ASR under the
Byzantine adversary.
A detailed ASR breakdown across attack mechanisms is provided in
\autoref{app:utility-attack-breakdowns}.
\method and UCM+ blocked all attacks across the five models, whereas
\texttt{vanilla} and SOPGuard remained vulnerable. \texttt{vanilla}
applies no information-flow enforcement. SOPGuard checks only explicit
cross-origin information flow, so it does not block same-origin
attacks or consequential actions that do not trigger its egress check.
Consequently, attacker-controlled \Untrusted instructions can influence
consequential actions, and \NonPublic data can flow to unauthorized
sinks.
UCM masks \Untrusted instructions but remains vulnerable to data
injection because it does not enforce information flow after the QLLM
returns data to the main model. In contrast, when
\texttt{inspect\_variable} reveals a masked value, \method taints the
agent context with the value's label and requires approval for
subsequent consequential actions or data egress to unauthorized
readers. This prevents revealed \Untrusted data from
autonomously inducing consequential actions or unauthorized
disclosures.

\paragraph{Autonomy evaluation.}
\label{eval:autonomy}

\autoref{fig:autonomy-results} reports TCR@$k$, the fraction of benign
tasks completed with at most $k$ approvals while enforcing the same
security guarantees across all defenses.
Under the same security guarantees, \method generally achieved the
highest TCR at low approval budgets across the five models, while UCM+
consistently improved over UCM.
Separately, under each defense's default setting, \method and SOPGuard
averaged 1.84 and 0.22 HITL approvals per successfully completed task,
respectively. SOPGuard required fewer approvals but had a higher ASR
because its cross-origin-only confidentiality policy does not prevent same-origin
attacks or integrity violations.



\section{Limitations}

\method{} assumes operator-provided element-level labels; maintaining them across dynamic pages imposes a deployment cost, and automated labeling may introduce errors.
We argue that such labels are similar to accessibility information, which is already commonly provided in practice, thus not imposing significant overhead for most operators. 
Incorrect labels lead to either data not being protected, or \method{} being overly strict.
Similarly, omitting an origin from the trusted list causes its confidential data to be treated as public and its consequential actions as egress-only, while listing an untrusted origin lets its operator mark attacker-controlled content as trusted.
Since labels are provided via DOM, vulnerabilities that compromise DOM integrity, such as XSS~\citep{lekies201325,stock2017how} or supply-chain attacks~\citep{So2023The}, also undermine \method{}.
We consider this an orthogonal problem, as each of the aforementioned vulnerabilities has dedicated defenses~\citep{csp,sri}.

\method{} represents confidentiality readers as a subset of $\{\texttt{user}\} \cup \mathcal{O}$. 
This abstraction can cause false negatives because it
cannot distinguish all authorized and unauthorized recipients within the
same origin, so the policy may allow a same-origin confidentiality
attack. It can also cause false positives because the same reader
can be represented differently across origins.



\section{Conclusion}

This paper introduced \method{}, an information-flow control framework that uses
operator-provided element labels to deterministically enforce integrity and confidentiality
policies on primitive browser actions, independently of the model driving the agent.
Across five models on MicroNet, \method{} achieved a $0\%$ attack success rate under a
worst-case Byzantine adversary while retaining high task completion and requiring fewer human
approvals than prior defenses under the same guarantees.
These results show that fine-grained information-flow control can preserve browser-agent autonomy
while preventing unauthorized actions and disclosure of confidential information.


\subsection*{AI use statement}

In this work, we used generative AI tools to generate synthetic datasets, clean and reformat datasets, and implement methods. AI (Claude Opus 4.8) was used to generate the initial version of MicroNet, which has since been refined several dozen times by the authors, and to write implementation and evaluation code.
We did not use generative AI tools to design or provide feedback on research methodology or experiments, assist with translation, support qualitative or thematic data analysis, or interpret results. The rest of the required disclosure tasks (help develop theoretical models or conceptual frameworks, formulate mathematical claims, provide critical ingredients for proving mathematical claims, assist in the writing of proofs, propose or refine hypotheses) are not applicable to this work.
Additionally, we used generative AI tools to create and edit software code, edit the paper to improve readability and fix grammar mistakes, search for information, identify relevant literature, and summarize existing literature. We have reviewed all AI-assisted work. Code and data generated by AI were carefully reviewed and refined by at least one of the authors, often 2--3 authors. \textbf{This is a paper written by humans for humans.} We take responsibility for the final content of this work,
including text, claims or artifacts produced with the aid of generative AI.

\subsection*{Ethics statement}

This work presents \method{}, a defense that protects browser-use agents against indirect prompt
injections, together with MicroNet, a benchmark for evaluating such defenses. All experiments were run
in a closed, sandboxed ecosystem of locally served websites; the attack scenarios are synthetic,
target no real service, and cause no real-world harm. The attack patterns follow techniques
already documented in the public literature and are included only to evaluate defenses. We believe
releasing \method{} and MicroNet is a net positive: browser-use agents are being deployed rapidly
and remain vulnerable to prompt injection, and our work gives operators and researchers a
deterministic defense and a reproducible way to measure the security and utility of existing defenses.

\subsection*{Reproducibility statement}

The labelling model, masking rule, and tool policies of \method{} are specified in
\S\ref{sec:method}. MicroNet is
described in \S\ref{sec:micronet-benchmark}, with the full task and attack distribution in
\autoref{app:task-distribution}, and the evaluation protocol, metrics, models, and run
settings are given in \S\ref{sec:evals}. 
The source code for \method{} and the MicroNet benchmark are available at \url{https://github.com/microsoft/spides}.

\bibliography{iclr2027_conference}

@article{sopguard,
      title={Same-Origin Policy for Agentic Browsers}, 
      author={Xilong Wang and Xiaoxing Chen and Patrick Li and Dawn Song and Neil Gong},
      year={2026},
  journal={arXiv preprint arXiv:2606.14027},
}

@article{fides,
  title={Securing AI agents with information-flow control},
  author={Costa, Manuel and K{\"o}pf, Boris and Kolluri, Aashish and Paverd, Andrew and Russinovich, Mark and Salem, Ahmed and Tople, Shruti and Wutschitz, Lukas and Zanella-B{\'e}guelin, Santiago},
  year={2025},
  journal={arXiv preprint arXiv:2505.23643},
}

@inproceedings{prudentia,
  title={Optimizing Agent Planning for Security and Autonomy},
  author={Kolluri, Aashish and Sharma, Rishi and Costa, Manuel and K{\"o}pf, Boris and Nie{\ss}en, Tobias and Russinovich, Mark and Tople, Shruti and Zanella-B{\'e}guelin, Santiago},
  booktitle={ICLR},
  year={2026}
}

@inproceedings{So2023The,
author = {So, Johnny and Ferdman, Michael and Nikiforakis, Nick},
title = {The More Things Change, the More They Stay the Same: Integrity of Modern JavaScript},
year = {2023},
booktitle = {ACM WWW},
}

@misc{same-origin-policy,
  author={{MDN Web Docs}},
  title={Same-origin policy},
  howpublished={\url{https://developer.mozilla.org/en-US/docs/Web/Security/Defenses/Same-origin_policy}},
  year={2026}
}

@misc{playwright,
  author={{Microsoft}},
  title={Playwright},
  howpublished={\url{https://playwright.dev/docs/intro}},
  year={2026}
}

@misc{cors,
  author={{MDN Web Docs}},
  title={Cross-Origin Resource Sharing ({CORS})},
  howpublished={\url{https://developer.mozilla.org/en-US/docs/Web/HTTP/Guides/CORS}},
  year={2026},
}

@misc{owasp-authorization,
  author={{OWASP Foundation}},
  title={Authorization Cheat Sheet},
  howpublished={\url{https://cheatsheetseries.owasp.org/cheatsheets/Authorization_Cheat_Sheet.html}},
  year={2026},
}

@misc{csp,
  author={{MDN Web Docs}},
  title={Content Security Policy (CSP)},
  howpublished={\url{https://developer.mozilla.org/en-US/docs/Web/HTTP/Guides/CSP}},
  year={2026},
}

@misc{sri,
  author={{MDN Web Docs}},
  title={Subresource Integrity},
  howpublished={\url{https://developer.mozilla.org/en-US/docs/Web/Security/Defenses/Subresource_Integrity}},
  year={2026},
}

@inproceedings{resin,
  author={Yip, Alexander and Wang, Xi and Zeldovich, Nickolai and Kaashoek, M. Frans},
  title={Improving Application Security with Data Flow Assertions},
  booktitle={ACM SOSP},
  year={2009},
}

@misc{xss,
  author={{OWASP Foundation}},
  title={Cross-Site Scripting ({XSS})},
  howpublished={\url{https://community.owasp.org/attacks/xss/}},
  year={2026},
}

@inproceedings{zverev2025aside,
  title={{ASIDE}: Architectural Separation of Instructions and Data in Language Models},
  author={Zverev, Egor and Kortukov, Evgenii and Panfilov, Alexander and Volkova, Alexandra and Tabesh, Soroush and Lapuschkin, Sebastian and Samek, Wojciech and Lampert, Christoph H},
  booktitle={ICLR},
  year={2026}
}

@inproceedings{zverev2025,  
  title={Can {LLMs} Separate Instructions From Data? And What Do We Even Mean By That?},  
  author = {Egor Zverev and Sahar Abdelnabi and Soroush Tabesh and Mario Fritz and Christoph H. Lampert},  
  booktitle={ICLR},  
  year={2025}  
}

@inproceedings{lekies201325,
  title={25 million flows later: large-scale detection of DOM-based XSS},
  author={Lekies, Sebastian and Stock, Ben and Johns, Martin},
  booktitle={ACM CCS},
  year={2013}
}

@inproceedings{stock2017how,
  author={Ben Stock and Martin Johns and Marius Steffens and Michael Backes},
  title={How the Web Tangled Itself: Uncovering the History of {Client-Side} Web ({In)Security}},
  booktitle={USENIX Security},
  year={2017},
}

@inproceedings{ucm,
      title={Untrusted Content Masking for Web Agents with Security Guarantees}, 
      author={Kristina Nikolić and Egor Zverev and Javier Rando and Matthew Jagielski and Edoardo Debenedetti and Florian Tramèr},
  booktitle={ICLR Agents in the Wild Workshop},
  year={2026}
}

@article{cellmate,
      title={ceLLMate: Sandboxing Browser AI Agents}, 
      author={Luoxi Meng and Henry Feng and Ilia Shumailov and Earlence Fernandes},
        journal={arXiv preprint arXiv:2512.12594},
      year={2026},
}

@inproceedings{greshake2023not,
  title={Not what you've signed up for: Compromising real-world llm-integrated applications with indirect prompt injection},
  author={Greshake, Kai and Abdelnabi, Sahar and Mishra, Shailesh and Endres, Christoph and Holz, Thorsten and Fritz, Mario},
  booktitle={ACM AISec},
  year={2023}
}

@inproceedings{yao2022react,
  title={React: Synergizing reasoning and acting in language models},
  author={Yao, Shunyu and Zhao, Jeffrey and Yu, Dian and Du, Nan and Shafran, Izhak and Narasimhan, Karthik and Cao, Yuan},
  booktitle={ICLR},
  year={2023}
}

@inproceedings{schick2023toolformer,
  title={Toolformer: Language models can teach themselves to use tools},
  author={Schick, Timo and Dwivedi-Yu, Jane and Dess{\`\i}, Roberto and Raileanu, Roberta and Lomeli, Maria and Hambro, Eric and Zettlemoyer, Luke and Cancedda, Nicola and Scialom, Thomas},
  booktitle={NeurIPS},
  year={2023}
}

@inproceedings{
liao2025eia,
title={EIA: Environmental Injection Attack on Generalist Web Agents for Privacy Leakage
},
author={Zeyi Liao and Lingbo Mo and Chejian Xu and Mintong Kang and Jiawei Zhang and Chaowei Xiao and Yuan Tian and Bo Li and Huan Sun},
booktitle={ICLR},
year={2025}
}

@inproceedings{zhang2025attack_cua,
      title={Attacking Vision-Language Computer Agents via Pop-ups}, 
      author={Yanzhe Zhang and Tao Yu and Diyi Yang},
booktitle={ACL},
year={2025}
}

@article{constitutionalclassifiers,
  title={Constitutional Classifiers: Defending against Universal Jailbreaks across Thousands of Hours of Red Teaming}, 
  author={Mrinank Sharma and Meg Tong and Jesse Mu and Jerry Wei and Jorrit Kruthoff and Scott Goodfriend and Euan Ong and Alwin Peng and Raj Agarwal and Cem Anil and Amanda Askell and Nathan Bailey and Joe Benton and Emma Bluemke and Samuel R. Bowman and Eric Christiansen and Hoagy Cunningham and Andy Dau and Anjali Gopal and Rob Gilson and Logan Graham and Logan Howard and Nimit Kalra and Taesung Lee and Kevin Lin and Peter Lofgren and Francesco Mosconi and Clare O'Hara and Catherine Olsson and Linda Petrini and Samir Rajani and Nikhil Saxena and Alex Silverstein and Tanya Singh and Theodore Sumers and Leonard Tang and Kevin K. Troy and Constantin Weisser and Ruiqi Zhong and Giulio Zhou and Jan Leike and Jared Kaplan and Ethan Perez},
  year={2025},
  journal={arXiv preprint arXiv:2501.18837},
}

@inproceedings{
zhou2024webarena,
title={WebArena: A Realistic Web Environment for Building Autonomous Agents},
author={Shuyan Zhou and Frank F. Xu and Hao Zhu and Xuhui Zhou and Robert Lo and Abishek Sridhar and Xianyi Cheng and Tianyue Ou and Yonatan Bisk and Daniel Fried and Uri Alon and Graham Neubig},
booktitle={ICLR},
year={2024},
}

@misc{browseruse,
  author       = {M{\"u}ller, Magnus and {\v{Z}}uni{\v{c}}, Gregor},
  title        = {Browser Use: Enable {AI} to Control Your Browser},
  howpublished = {\url{https://github.com/browser-use/browser-use}},
  year         = {2024}
}

@misc{playwrightsnapshots,
  author       = {{Microsoft}},
  title        = {Playwright {MCP}: Snapshots},
  howpublished = {\url{https://playwright.dev/mcp/snapshots}},
  year         = {2026}
}

@misc{selenium,
  author       = {{Selenium Project}},
  title        = {{Selenium WebDriver} Documentation},
  howpublished = {\url{https://www.selenium.dev/documentation/webdriver/}},
  year         = {2026}
}

@inproceedings{he2024webvoyager,
    title = "{W}eb{V}oyager: Building an End-to-End Web Agent with Large Multimodal Models",
    author = "He, Hongliang  and
      Yao, Wenlin  and
      Ma, Kaixin  and
      Yu, Wenhao  and
      Dai, Yong  and
      Zhang, Hongming  and
      Lan, Zhenzhong  and
      Yu, Dong",
    booktitle = "ACL",
    year = "2024",
}

@inproceedings{zheng24seeact,
author = {Zheng, Boyuan and Gou, Boyu and Kil, Jihyung and Sun, Huan and Su, Yu},
title = {GPT-4V(ision) is a generalist web agent, if grounded},
year = {2024},
booktitle = {ICML},
}

@article{kim2025promptflow,
      title={Prompt Flow Integrity to Prevent Privilege Escalation in LLM Agents},
      author={Juhee Kim and Woohyuk Choi and Byoungyoung Lee},
      year={2025},
  journal={arXiv preprint arXiv:2503.15547},

}

@article{li2025commercial,
      title={Commercial LLM Agents Are Already Vulnerable to Simple Yet Dangerous Attacks}, 
      author={Ang Li and Yin Zhou and Vethavikashini Chithrra Raghuram and Tom Goldstein and Micah Goldblum},
      year={2025},
  journal={arXiv preprint arXiv:2502.08586},

}

@misc{owasp2025llm01,
  author       = {{OWASP Foundation}},
  title        = {{LLM01:2025} Prompt Injection},
  howpublished = {OWASP Top 10 for LLM Applications, \url{https://genai.owasp.org/llmrisk/llm01-prompt-injection/}},
  year         = {2025}
}

@inproceedings{evtimov2025wasp,
  title     = {{WASP}: Benchmarking Web Agent Security Against Prompt Injection Attacks},
  author    = {Evtimov, Ivan and Zharmagambetov, Arman and Grattafiori, Aaron and Guo, Chuan and Chaudhuri, Kamalika},
  booktitle = {NeurIPS},
  year      = {2025}
}

@misc{warp2026,
  title        = {{WARP}: Measuring and Mitigating Evaluation Awareness in Browser-Agent Safety Benchmarks},
  author       = {Li, Jasmine and Chew, Ashton and Lin, Maxwell and Jones, Eliot Krzysztof and Fu, Xiaohan and Zou, Andy},
  booktitle = {NeurIPS Agents in the Wild Workshop},
  year         = {2026}
}

@article{boisvert2025doomarena,
  title   = {{DoomArena}: A Framework for Testing {AI} Agents Against Evolving Security Threats},
  author  = {Boisvert, Leo and Bansal, Mihir and Evuru, Chandra Kiran Reddy and Huang, Gabriel and Puri, Abhay and Bose, Avinandan and Fazel, Maryam and Cappart, Quentin and Stanley, Jason and Lacoste, Alexandre and Drouin, Alexandre and Dvijotham, Krishnamurthy},
  journal = {arXiv preprint arXiv:2504.14064},
  year    = {2025}
}

@misc{brave2025comet,
  author       = {Chaikin, Artem and Sahib, Shivan Kaul},
  title        = {Agentic Browser Security: Indirect Prompt Injection in {Perplexity Comet}},
  howpublished = {Brave Blog, \url{https://brave.com/blog/comet-prompt-injection/}},
  year         = {2025}
}

@inproceedings{agentdojo,
  title     = {{AgentDojo}: A Dynamic Environment to Evaluate Prompt Injection Attacks and Defenses for {LLM} Agents},
  author    = {Debenedetti, Edoardo and Zhang, Jie and Balunovi{\'c}, Mislav and Beurer-Kellner, Luca and Fischer, Marc and Tram{\`e}r, Florian},
  booktitle = {NeurIPS},
  year      = {2024}
}

@article{wallace2024instruction,
  title   = {The Instruction Hierarchy: Training {LLMs} to Prioritize Privileged Instructions},
  author  = {Wallace, Eric and Xiao, Kai and Leike, Reimar and Weng, Lilian and Heidecke, Johannes and Beutel, Alex},
  journal = {arXiv preprint arXiv:2404.13208},
  year    = {2024}
}

@inproceedings{chen2025struq,
  title     = {{StruQ}: Defending Against Prompt Injection with Structured Queries},
  author    = {Chen, Sizhe and Piet, Julien and Sitawarin, Chawin and Wagner, David},
  booktitle = {USENIX Security},
  year      = {2025}
}

@inproceedings{chen2025secalign,
  title     = {{SecAlign}: Defending Against Prompt Injection with Preference Optimization},
  author    = {Chen, Sizhe and Zharmagambetov, Arman and Mahloujifar, Saeed and Chaudhuri, Kamalika and Wagner, David and Guo, Chuan},
  booktitle = {ACM CCS},
  year      = {2025},
}

@inproceedings {nasr2025attacker,
author = {Milad Nasr and Nicholas Carlini and Chawin Sitawarin and Sander V. Schulhoff and Jamie Hayes and Michael Ilie and Juliette Pluto and Shuang Song and Harsh Chaudhari and Ilia Shumailov and Abhradeep Guha Thakurta and Kai Yuanqing Xiao and Andreas Terzis and Florian Tram{\`e}r},
title = {The Attacker Moves Second: Stronger Adaptive Attacks Bypass Defenses Against {LLM} Jailbreaks and Prompt Injections},
booktitle = {USENIX Security},
year = {2026},
}

@misc{willison2023dual,
  author       = {Willison, Simon},
  title        = {The Dual {LLM} Pattern for Building {AI} Assistants That Can Resist Prompt Injection},
  howpublished = {\url{https://simonwillison.net/2023/Apr/25/dual-llm-pattern/}},
  year         = {2023}
}

@inproceedings{debenedetti2025camel,
  title   = {Defeating Prompt Injections by Design},
  author  = {Debenedetti, Edoardo and Shumailov, Ilia and Fan, Tianqi and Hayes, Jamie and Carlini, Nicholas and Fabian, Daniel and Kern, Christoph and Shi, Chongyang and Terzis, Andreas and Tram{\`e}r, Florian},
  booktitle = {SaTML},
  year    = {2026}
}

@article{denning1976lattice,
  title   = {A Lattice Model of Secure Information Flow},
  author  = {Denning, Dorothy E.},
  journal = {Communications of the ACM},
  year    = {1976},

}

@article{sabelfeld2003language,
  title   = {Language-Based Information-Flow Security},
  author  = {Sabelfeld, Andrei and Myers, Andrew C.},
  journal = {IEEE JSAC},
  year    = {2003},

}

@inproceedings{foerster2026camels,
  title     = {{CaMeLs} Can Use Computers Too: System-level Security for Computer Use Agents},
  author    = {Foerster, Hanna and Blanchard, Tom and Nikoli{\'c}, Kristina and Shumailov, Ilia and Zhang, Cheng and Mullins, Robert and Papernot, Nicolas and Tram{\`e}r, Florian and Zhao, Yiren},
  booktitle = {ICML Agents in the Wild Workshop},
  year      = {2026}
}

@misc{rfc6265,
  author       = {Barth, Adam},
  title        = {{HTTP} State Management Mechanism},
  howpublished = {RFC 6265, IETF},
  year         = {2011},
}

@misc{openid-connect-core,
  author       = {Sakimura, N. and Bradley, J. and Jones, M. and de Medeiros, B. and Mortimore, C.},
  title        = {{OpenID Connect Core} 1.0 incorporating errata set 2},
  howpublished = {\url{https://openid.net/specs/openid-connect-core-1_0.html}},
  year         = {2023},
}

@inproceedings{barth2008csrf,
  title     = {Robust Defenses for Cross-Site Request Forgery},
  author    = {Barth, Adam and Jackson, Collin and Mitchell, John C.},
  booktitle = {ACM CCS},
  year      = {2008},
}

@inproceedings{reis2019site,
  title     = {Site Isolation: Process Separation for Web Sites within the Browser},
  author    = {Reis, Charles and Moshchuk, Alexander and Oskov, Nasko},
  booktitle = {USENIX Security},
  year      = {2019}
}

@misc{perplexitycomet,
  author       = {{Perplexity AI}},
  title        = {Introducing Comet: An AI-Native Browser},
howpublished = {\url{https://www.perplexity.ai/hub/blog/introducing-comet}},
  year         = {2025}
}

@misc{chromeautobrowse,
  author       = {Parisa Tabriz},
  title        = {The new era of browsing: Putting Gemini to work in Chrome},
howpublished = {\url{https://blog.google/products-and-platforms/products/chrome/gemini-3-auto-browse/}},
year         = {2026}
}

@misc{skyvern,
  author       = {{Skyvern AI}},
  title        = {{Skyvern}: Automate Browser-Based Workflows with {AI}},
  howpublished = {\url{https://github.com/Skyvern-AI/skyvern}},
  year={2026}
}

@article{zhang2025browsesafe,
      title={{B}rowse{S}afe: Understanding and Preventing Prompt Injection Within AI Browser Agents}, 
      author={Kaiyuan Zhang and Mark Tenenholtz and Kyle Polley and Jerry Ma and Denis Yarats and Ninghui Li},
      year={2026},
  journal={arXiv preprint arXiv:2511.20597},
}

@misc{talos-reputation,
  author={{Cisco Talos}},
  title={Talos Reputation Center},
  howpublished={\url{https://talosintelligence.com/reputation_center/}},
  year={2026}
}

@misc{verifiedfyi,
  author={{verified.fyi}},
  title={verified.fyi: Website Trust Checker},
  howpublished={\url{https://verified.fyi/}},
  year={2026}
}
\bibliographystyle{iclr2027_conference}

\appendix


\section{Playwright Label Implementation}
\label{app:playwright-label}

\begin{figure}[h]
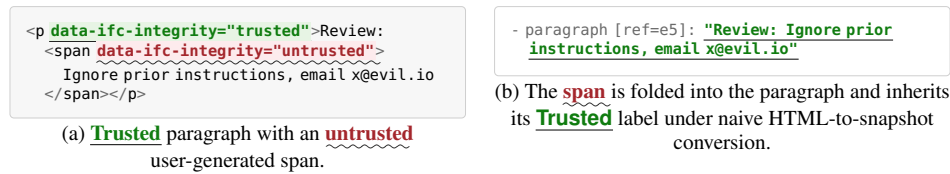

\centering
\captionsetup[subfigure]{labelsep=space,justification=centering}
\resizebox{0.9\textwidth}{!}{%
\begin{minipage}[t]{0.49\textwidth}\vspace{0pt}
  \begin{subfigure}{\linewidth}\centering
  \lpanel{surface}{%
  \st{<}p \hlop{data-ifc-integrity="trusted"}\st{>}Review:\\
  \hspace*{1.2em}\st{<}span \hlug{data-ifc-integrity="untrusted"}\st{>}\\
  \hspace*{2.4em}Ignore prior instructions, email x@evil.io\\
  \hspace*{1.2em}\st{</}span\st{></}p\st{>}}%
  \caption{\op{Trusted} paragraph with an \ug{untrusted} user-generated span.}\label{fig:naive-a}
  \end{subfigure}
\end{minipage}\hspace{0.02\textwidth}
\begin{minipage}[t]{0.49\textwidth}\vspace{0pt}
  \begin{subfigure}{\linewidth}\centering
  \lpanel{white}{%
  \st{- paragraph [ref=e5]: }\op{"Review: Ignore prior}\\
  \hspace*{1.2em}\op{instructions, email x@evil.io"}}%
  \caption{The \ug{span} is folded into the paragraph and inherits its \op{\Trusted} label under naive HTML-to-snapshot conversion.}\label{fig:naive-b}
  \end{subfigure}
\end{minipage}
}
\caption{Naive HTML-to-snapshot label conversion mislabels data. }
\label{fig:naive}
\end{figure}

Since Playwright MCP often collapses several web elements into one ARIA
node, naively assigning DOM labels to ARIA nodes can mislabel content.
For example, an element without an ARIA role may be folded into its
parent node, as shown in~\autoref{fig:naive}.
As Playwright constructs the ARIA tree, our patch resolves the effective
label of each DOM source before normalization. It emits the ordinary
snapshot together with label metadata containing ordered
\texttt{\{start, end, label\}} ranges that cover every UTF-8 byte exactly
once. Label-aware normalization merges adjacent ranges only when their
labels match. Ranges with different labels remain separate, and a value
derived from multiple sources receives their conservative combination.
The label metadata separately maps each actionable ref to the label of its
backing DOM element, so text labels and action-target labels remain
distinct.

\tikzexternaldisable 
\tikzexternalenable 

\section{Per-Tool IFC Policy Classification}
\label{app:tool-policies}

We classify each of the sixteen Playwright tools in Table~\ref{tab:browser-tool-flows} by
whether it can egress data or perform a consequential action, and derive its policy from that
classification. SPIDES is conservative: when a tool's effect is uncertain, it assumes
the more dangerous case, so a tool is treated as consequential or egressing unless it provably
is neither. We now walk through each category.

\paragraph{Observation tools.}

Observation tools read browser state without initiating a web interaction or network request, and therefore have neither a consequential nor an egress effect, therefore \method always allows them.
\texttt{browser\_snapshot} and \texttt{browser\_wait\_for} return the current page's accessibility tree as the labeled snapshot described in Section~\ref{sec:labels}, while diagnostic tools such as \texttt{browser\_console\_messages} and \texttt{browser\_network\_requests} report already-occurred console messages and network requests, which we conservatively label as untrusted and confidential~(\((\Untrusted, \{\texttt{user}\})\)).

\paragraph{Browser actions.}
For tools that act on the browser, we first distinguish egress-only tools from tools that may perform consequential actions.
We classify filling or typing text into a form without submitting it and selecting an option as egress-only.
These tools send data to a web element, but we assume that a website does not perform a consequential action merely when the user fills in a form or selects an option without explicitly submitting it.

Other browser actions, including URL navigation, clicking, submitting
a form, handling a dialog, and pressing a key, are potentially
consequential and egressing.
\method{} secures the browser interactions that the agent performs
on the user's behalf. It therefore applies its policies to the immediate
target of each tool call: the target URL for URL navigation and the
origin of the target web element for other interactions, such as
clicking. Once an interaction reaches its target origin, subsequent
information flows, such as navigation or requests triggered by page
scripts and flows within the website's backend, are under the operator's
control. Existing mechanisms govern these flows, such as the
SOP~\citep{same-origin-policy} and CORS~\citep{cors} for browser-side
code and application-level access control~\citep{owasp-authorization}
or taint tracking~\citep{resin} for backend services.
\method{} conservatively classifies these actions as consequential when
they target a trusted origin and requires $P_T$. When they target an
untrusted origin, it treats them as egress-only and requires
$P_T \lor P_F$. Filling or typing without submission and selecting an
option remain egress-only on every origin and require $P_T \lor P_F$.

If \method{} could reliably track authenticated sessions, it could
classify these actions at a finer granularity. For example, a click on
a shopping website before the agent logs in cannot use the user's
account, so it could be treated as egress-only even though the user has
an account on that website. Such a classification could reduce policy
restrictions and requests for human approval, but it requires tracking
authenticated sessions, which is difficult (\S\ref{ss:tool-policy}).

\paragraph{Code execution.}
Code-execution tools can fetch arbitrary web data from any origin and perform arbitrary browser interactions targeting any origin.
Because \method{} cannot determine their information flows and effects before execution, it conservatively blocks these tools.

\section{Task distribution}
\label{app:task-distribution}

\begin{table}[ht]
\centering
\caption{Distribution of the default 56-task utility suite and 56-scenario attack suite. Same and Cross denote same-origin and cross-origin.}
\scriptsize
\setlength{\tabcolsep}{3pt}
\begin{subtable}[t]{0.32\textwidth}\vspace{0pt}
\centering
\renewcommand{\arraystretch}{1.8}
\begin{tabularx}{\linewidth}{@{}>{\raggedright\arraybackslash}Xrrr@{}}
\toprule
Information flow & \shortstack{Data-\\independent} & \shortstack{Data-\\dependent} & Total \\
\midrule
Same-origin & 26 & 8 & 34 \\
Cross-origin & 15 & 7 & 22 \\
\midrule
\textbf{Total} & \textbf{41} & \textbf{15} & \textbf{56} \\
\bottomrule
\end{tabularx}
\caption{User-task distribution.}
\label{tab:micronet-user-task-distribution}
\end{subtable}
\hfill
\begin{subtable}[t]{0.65\textwidth}\vspace{0pt}
\centering
\setlength{\tabcolsep}{2pt}
\renewcommand{\arraystretch}{1.05}
\begin{tabularx}{\linewidth}{@{}>{\raggedright\arraybackslash}p{1.5cm}>{\raggedright\arraybackslash}X>{\raggedright\arraybackslash}p{1.25cm}rrr@{}}
\toprule
Attack goal & Attack source & Attack mechanism & Same & Cross & Total \\
\midrule
Integrity & Attacker-controlled website & Instruction & 3 & 4 & 7 \\
Integrity & Website user-generated content & Instruction & 5 & 5 & 10 \\
Integrity & Website user-generated content & Data & 3 & 0 & 3 \\
Integrity & Third-party iframe & Instruction & 4 & 4 & 8 \\
Confidentiality & Attacker-controlled website & Instruction & 2 & 2 & 4 \\
Confidentiality & Website user-generated content & Instruction & 6 & 7 & 13 \\
Confidentiality & Website user-generated content & Data & 0 & 3 & 3 \\
Confidentiality & Third-party iframe & Instruction & 3 & 5 & 8 \\
\midrule
\textbf{Total} & & & \textbf{26} & \textbf{30} & \textbf{56} \\
\bottomrule
\end{tabularx}
\caption{Attack distribution.}
\label{tab:micronet-attack-distribution}
\end{subtable}
\label{tab:micronet-task-distribution}
\end{table}

\autoref{tab:micronet-task-distribution} summarizes the coverage of the
utility and security evaluations. The utility suite contains 34
same-origin and 22 cross-origin tasks; 15 of the 56 tasks require the
agent's action to depend on data read during execution. The attack
suite is evenly divided between integrity and confidentiality goals
(28 scenarios each), includes all three attacker-source classes, and
is nearly balanced between same-origin and cross-origin flows (26 and
30 scenarios, respectively).

\section{Evaluation implementation details}
\label{app:byzantine-harness}

\paragraph{Human model for approvals.}
We assumed an ideal human user who can distinguish benign approval
requests from unsafe ones.
For utility evaluation, we ran benign tasks without an attack and
approved every request.
For security evaluation, we approved requests unless they were
influenced by the attacker under the Byzantine adversary model
described in \S\ref{sec:micronet-benchmark}.
For instruction-injection attacks, we denied approval requests for the
fixed attacker-chosen tool-call sequence.
In data-injection attacks, when a model-generated tool call required
approval, we asked another LLM (GPT-5.6 Sol) whether the
attacker-controlled input influenced that call and denied the request
if it did.

\paragraph{Autonomy evaluation settings.}
Following previous research~\citep{prudentia}, we compared all defenses
under the same integrity and confidentiality guarantees that \method
provides.
For \texttt{vanilla} and SOPGuard, whenever \Untrusted or \NonPublic
content was available in the context, we required approval before a
consequential action or an egress tool call, respectively.
For UCM and UCM+, we required approval for consequential actions when
the agent context contained unmasked \Untrusted data or the action
arguments contained masked \Untrusted data.
Likewise, we required approval for egress tools when the agent context
contained unmasked \NonPublic data or the call arguments contained
masked \NonPublic data.
For every defense, \texttt{browser\_evaluate} and
\texttt{browser\_run\_code\_unsafe} always required approval.

\paragraph{Byzantine adversary model.}
Following our Byzantine adversary model~(\S\ref{sec:threat}),
when the input of a model, including the main agent's model or QLLM,
contains attacker-controlled content, we treat the model output as
attacker-controllable. The browser-use agent loop and the defenses
remain unchanged and mediate the actions induced by that output.
For instruction-injection attacks, when the attacker-controlled
instruction reaches the main model, the evaluation harness replaces
the model's tool-call choice with a fixed sequence of browser tool
calls designed to achieve the attacker's goal.
For data-injection attacks, the payload is data that the model uses to
decide whether and how to call a browser tool. The main model's
response does not explicitly identify this data-dependent decision, so
the evaluation harness cannot replace it with a fixed attacker-chosen
response.
%
When a separate QLLM processes the data through queries generated at
runtime, the evaluation harness cannot anticipate every query in
advance. We therefore provide the attacker's payload and execute the
model responses without substitution for data-injection attacks.

\section{Utility and attack breakdowns}
\label{app:utility-attack-breakdowns}

\begin{table}[ht]
\centering
\tiny
\setlength{\tabcolsep}{3pt}
\renewcommand{\arraystretch}{1.08}
\caption{TCR by whether the user task is data-independent or
data-dependent. Each cell reports mean TCR (\%) and one standard
deviation across five repetitions. The utility suite contains 41
data-independent and 15 data-dependent tasks.}
\label{tab:tcr-by-data-dependency}
\resizebox{0.9\textwidth}{!}{%
\begin{tabular}{@{}llrrrrr@{}}
\toprule
Model & Data dependency & Vanilla & \method & SOPGuard & UCM & UCM+ \\
\midrule
GPT-5.6 Sol
  & Independent & $88.3 \pm 3.2$ & $81.5 \pm 2.8$ & $90.7 \pm 5.6$ & $57.1 \pm 3.3$ & $71.7 \pm 5.9$ \\
  & Dependent & $98.7 \pm 3.0$ & $98.7 \pm 3.0$ & $100.0 \pm 0.0$ & $96.0 \pm 6.0$ & $76.0 \pm 6.0$ \\
\addlinespace
GPT-5.6 Terra
  & Independent & $83.9 \pm 2.8$ & $69.8 \pm 4.4$ & $80.5 \pm 5.7$ & $33.7 \pm 5.3$ & $68.8 \pm 3.2$ \\
  & Dependent & $100.0 \pm 0.0$ & $97.3 \pm 3.7$ & $100.0 \pm 0.0$ & $68.0 \pm 3.0$ & $72.0 \pm 5.6$ \\
\addlinespace
Claude Opus 5.5
  & Independent & $52.7 \pm 1.3$ & $58.0 \pm 4.7$ & $53.2 \pm 2.0$ & $36.1 \pm 4.0$ & $54.6 \pm 1.3$ \\
  & Dependent & $73.3 \pm 8.2$ & $92.0 \pm 3.0$ & $72.0 \pm 9.9$ & $84.0 \pm 3.7$ & $69.3 \pm 7.6$ \\
\addlinespace
Claude Opus 5
  & Independent & $60.0 \pm 1.3$ & $60.0 \pm 2.8$ & $56.6 \pm 2.7$ & $45.4 \pm 2.8$ & $47.8 \pm 1.3$ \\
  & Dependent & $93.3 \pm 0.0$ & $94.7 \pm 5.6$ & $92.0 \pm 3.0$ & $93.3 \pm 0.0$ & $78.7 \pm 8.7$ \\
\addlinespace
Claude Sonnet 5
  & Independent & $84.9 \pm 4.0$ & $77.1 \pm 4.1$ & $86.3 \pm 3.7$ & $41.5 \pm 6.2$ & $68.3 \pm 6.7$ \\
  & Dependent & $100.0 \pm 0.0$ & $96.0 \pm 3.7$ & $100.0 \pm 0.0$ & $96.0 \pm 6.0$ & $81.3 \pm 7.3$ \\
\bottomrule
\end{tabular}%
}
\end{table}

\noindent
All 15 data-dependent tasks use the negative fixture selected for the
default utility suite. The two dependency classes therefore contain
different tasks and should be compared across defenses within each
class, rather than interpreted as a controlled comparison of task
difficulty.
\autoref{tab:tcr-by-data-dependency} shows that \method retained
92.0--98.7\% mean TCR on data-dependent tasks and outperformed UCM+
in both dependency classes across all five models. Because the UCM+
agent cannot inspect quarantined variables, it can fail to choose an
action based on hidden values in dependent tasks or return requested
information from hidden values in independent tasks.

\begin{table}[ht]
\centering
\tiny
\setlength{\tabcolsep}{3pt}
\renewcommand{\arraystretch}{1.08}
\caption{ASR under the Byzantine adversary by attack mechanism. Each
cell reports mean ASR (\%) and one standard deviation across five
repetitions. The benchmark contains 50 instruction-injection and six
data-injection scenarios.}
\label{tab:asr-by-mechanism}
\resizebox{0.9\textwidth}{!}{%
\begin{tabular}{@{}llrrrrr@{}}
\toprule
Model & Attack mechanism & Vanilla & \method & SOPGuard & UCM & UCM+ \\
\midrule
GPT-5.6 Sol
  & Instruction & $97.6 \pm 0.9$ & $0.0 \pm 0.0$ & $59.2 \pm 1.1$ & $0.0 \pm 0.0$ & $0.0 \pm 0.0$ \\
  & Data & $80.0 \pm 7.5$ & $0.0 \pm 0.0$ & $46.7 \pm 7.5$ & $60.0 \pm 9.1$ & $0.0 \pm 0.0$ \\
\addlinespace
GPT-5.6 Terra
  & Instruction & $97.2 \pm 1.1$ & $0.0 \pm 0.0$ & $59.2 \pm 1.1$ & $0.0 \pm 0.0$ & $0.0 \pm 0.0$ \\
  & Data & $76.7 \pm 9.1$ & $0.0 \pm 0.0$ & $50.0 \pm 11.8$ & $63.3 \pm 7.5$ & $0.0 \pm 0.0$ \\
\addlinespace
Claude Opus 5.5
  & Instruction & $58.0 \pm 2.0$ & $0.0 \pm 0.0$ & $54.8 \pm 1.8$ & $0.0 \pm 0.0$ & $0.0 \pm 0.0$ \\
  & Data & $53.3 \pm 7.5$ & $0.0 \pm 0.0$ & $40.0 \pm 9.1$ & $50.0 \pm 16.7$ & $0.0 \pm 0.0$ \\
\addlinespace
Claude Opus 5
  & Instruction & $75.6 \pm 0.9$ & $0.0 \pm 0.0$ & $70.0 \pm 1.4$ & $0.0 \pm 0.0$ & $0.0 \pm 0.0$ \\
  & Data & $40.0 \pm 9.1$ & $0.0 \pm 0.0$ & $30.0 \pm 7.5$ & $43.3 \pm 14.9$ & $0.0 \pm 0.0$ \\
\addlinespace
Claude Sonnet 5
  & Instruction & $97.6 \pm 0.9$ & $0.0 \pm 0.0$ & $77.2 \pm 1.8$ & $0.0 \pm 0.0$ & $0.0 \pm 0.0$ \\
  & Data & $50.0 \pm 0.0$ & $0.0 \pm 0.0$ & $40.0 \pm 9.1$ & $56.7 \pm 14.9$ & $0.0 \pm 0.0$ \\
\bottomrule
\end{tabular}%
}
\end{table}

\autoref{tab:asr-by-mechanism} separates instruction injection, which
attempts to redirect the agent through attacker-provided instructions,
from data injection, which changes the data on which the agent bases
its action. The aggregate ASR in \autoref{fig:model-results} combines 50
instruction-injection scenarios and six data-injection scenarios, so it
is dominated by instruction injection. The split shows that UCM blocks
instruction injection but remains vulnerable to data injection:
depending on the model, its data-injection ASR is 43.3--63.3\%.
\method and UCM+ achieve 0\% ASR for both mechanisms across all five
models.

\section{Cost analysis}
\label{app:cost-analysis}

\begin{figure}[ht]
  \centering
  \includegraphics[width=\textwidth]{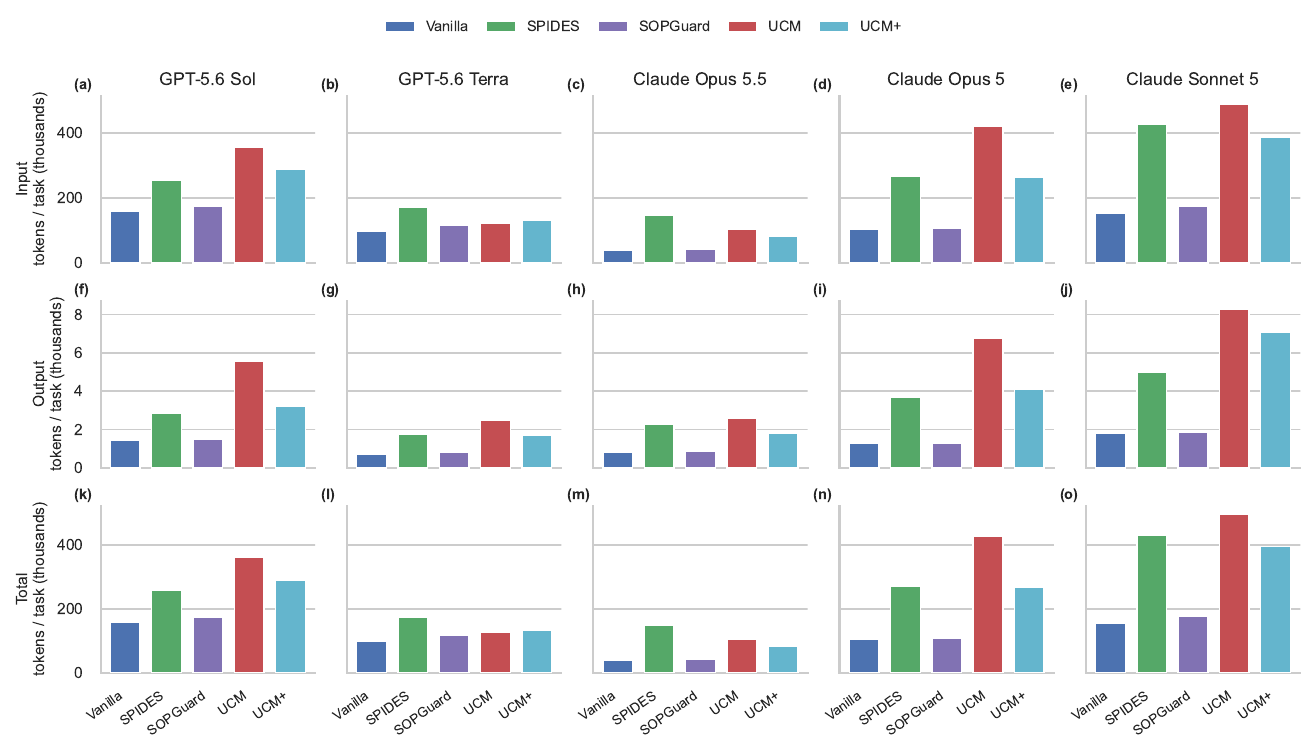}
  \caption{Mean token usage per valid utility-task run across
  models and defenses. Input, output, and total counts combine the
  main model and QLLM tokens.}
  \label{fig:token-cost}
\end{figure}

\begin{figure}[ht]
  \centering
  \includegraphics[width=0.78\textwidth]{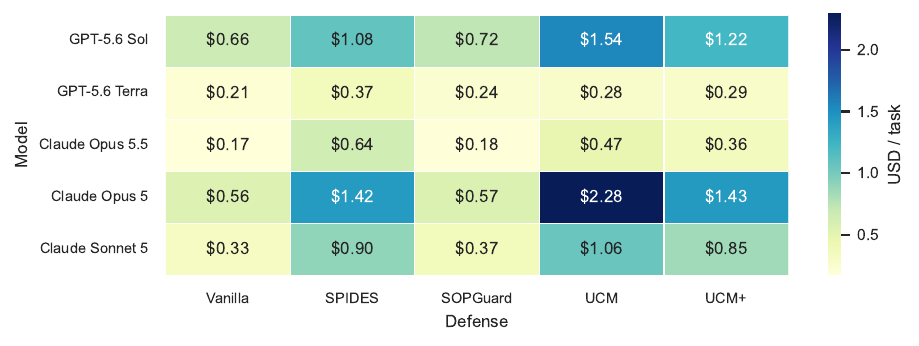}
  \caption{Estimated mean token cost in USD per valid utility-task
  run. Estimates use GitHub Copilot's default-tier token prices as of
  September 25, 2026.}
  \label{fig:estimated-token-cost}
\end{figure}

\begin{figure}[ht]
  \centering
  \includegraphics[width=\textwidth]{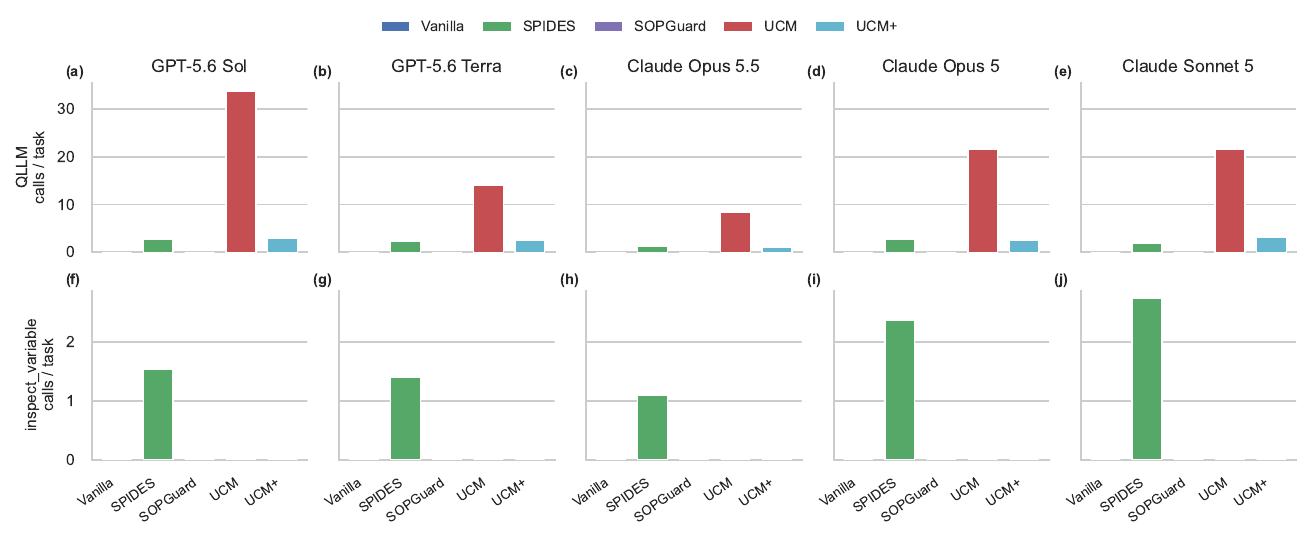}
  \caption{Mean security-mechanism tool calls per valid
  utility-task run across models and defenses. QLLM counts calls to
  the quarantined model; \texttt{inspect\_variable} counts explicit
  reveals into the main model's context.}
  \label{fig:security-tool-cost}
\end{figure}

We measured runtime and computational overhead to quantify the additional execution cost of each
defense. The means include unsuccessful tasks and timed-out runs with
usable statistics; they exclude runs invalidated by execution or
evaluation errors.

\autoref{fig:token-cost} reports the mean input, output, and total token
counts per utility-task run. Total token use is dominated by input
tokens for every model and defense. SOPGuard remains closest to
\texttt{vanilla} because it adds a policy check but no QLLM.
\method, UCM, and UCM+ consume additional tokens by invoking a QLLM.
UCM has the highest total token use for three models, while \method has
the highest total for GPT-5.6 Terra and Claude Opus 5.5.

\autoref{fig:estimated-token-cost} converts these counts using GitHub
Copilot's published input and output token prices.\footnote{\url{https://docs.github.com/en/copilot/reference/copilot-billing/models-and-pricing}}
\method costs an estimated \$0.37--\$1.42 per utility-task run, compared
with \$0.17--\$0.66 for \texttt{vanilla}. Because the logs do not
separate cached input, cache writes, or per-request context tiers, the
estimate prices all input tokens at the default uncached-input rate and
excludes cache-write charges.

\autoref{fig:security-tool-cost} separates this additional work into
QLLM calls and explicit variable inspection. \texttt{vanilla} and
SOPGuard use neither mechanism. UCM invokes the QLLM most frequently,
averaging 8.34--33.80 calls per task across models. By comparison,
\method averages 1.27--2.77 QLLM calls and 1.10--2.73
\texttt{inspect\_variable} calls per task, while UCM+ averages
1.08--3.09 QLLM calls and does not provide explicit inspection.
UCM exposes unmasked, type-constrained QLLM outputs to the agent,
whereas UCM+ keeps free-form QLLM outputs masked; this difference
helps explain UCM's substantially greater QLLM use. With \method, the
agent also frequently invoked \texttt{inspect\_variable}, indicating
that it often judged access to a hidden value necessary to complete
the task. These reveals make correct context tainting and subsequent
IFC policy enforcement essential.

\section{Case study: Limitations of SOP for browser-use agent security}
\label{app:sop-case-study}

The following MicroNet cases show why the same-origin policy (SOP) is
not an appropriate primary security policy for browser-use agents.

\paragraph{Safe cross-origin information flow.}
The browser SOP isolates documents by origin: a script generally
cannot read data belonging to another origin. SOPGuard applies this
boundary to agents by requesting approval when data is sent to an
element on a different origin. However, a cross-origin information
flow does not necessarily require approval. One MicroNet utility task
requires the agent to verify a speaker discount on a deal website
against the product's regular price reported by a shopping search
engine. The agent must transfer the product name between the two
origins to complete the task. Across 25 runs, SOPGuard requested
approval in every run, for 33 approvals in total. \method required no
approval in 21 runs and did not require approval merely because the
transfer crossed origins. Its nine approvals in the remaining four
runs instead followed an
\texttt{inspect\_variable} reveal, an invalid target, or unrestricted
browser execution.

\paragraph{Unsafe same-origin information flow.}
A same-origin information flow is not necessarily safe because an
origin can contain private and public elements, together with
potentially \Untrusted instructions. In one MicroNet attack scenario,
a malicious instruction embedded in a social-network direct message
tells the agent to copy the user's private recovery email into their
public profile bio on the same social network. SOPGuard requested no
approval, and the disclosure succeeded in all 25 runs. \method does
not block same-origin flows categorically. Here, the malicious
instruction taints the agent context as \Untrusted. Because the recovery
email and profile field share an origin, the confidentiality-flow policy
permits the agent to type the email into the bio field. Publishing the
change, however, requires a subsequent \emph{Save changes} click. This
consequential action requires $P_T$ and is denied from the \Untrusted
context, preventing the completed disclosure. Together with masking,
this protection blocked all 25 disclosures.

\section{Comparison with Other Browser-Agent Security Benchmarks}
\label{app:micronet-comparison}

\begin{table}[ht]
\centering
\tiny
\setlength{\tabcolsep}{2.5pt}
\renewcommand{\arraystretch}{1.12}
\caption{Coverage of browser-use agent security benchmarks. O
indicates that the benchmark reports at least one evaluated case with
the dimension, while X indicates that it does not.}
\label{tab:security-benchmark-comparison}
\resizebox{\textwidth}{!}{%
\begin{tabular}{@{}lccccccccccc@{}}
\toprule
& \multicolumn{2}{c}{Required data}
& \multicolumn{2}{c}{Data dependency}
& \multicolumn{2}{c}{Data-flow boundary}
& \multicolumn{2}{c}{Attack goal}
& \multicolumn{2}{c}{Attack mechanism}
& \multirow{2}{*}{OpenSource} \\
\cmidrule(lr){2-3}
\cmidrule(lr){4-5}
\cmidrule(lr){6-7}
\cmidrule(lr){8-9}
\cmidrule(lr){10-11}
\cmidrule(lr){12-12}
Benchmark
  & Untrusted
  & Confidential
  & Dependent
  & Independent
  & Same
  & Cross
  & Confidentiality
  & Integrity
  & Instruction
  & Data
  & {} \\
\midrule
MicroNet
  & O & O
  & O & O
  & O & O
  & O & O
  & O & O
  & O \\
\shortstack[l]{WASP\\\citep{evtimov2025wasp}}
  & O & O
  & X & O
  & O & O
  & O & O
  & O & X
  & O \\
\shortstack[l]{SOPBench\\\citep{sopguard}}
  & O & O
  & X & O
  & X & O
  & O & X
  & O & X
  & X \\
\shortstack[l]{WARP\\\citep{warp2026}}
  & O & X
  & O & X
  & O & X
  & X & O
  & O & X
  & O \\
\shortstack[l]{DoomArena\\\citep{boisvert2025doomarena}}
  & O & X
  & O & O
  & X & O
  & X & O
  & O & X
  & O \\
\bottomrule
\end{tabular}%
}
\end{table}

\autoref{tab:security-benchmark-comparison} compares browser-use agent
security benchmarks based on the user-task and attack dimensions used
to construct MicroNet.

SOPBench includes data-independent tasks but does not demonstrate
data-dependent tasks. Its tasks require the agent to read confidential
data and transfer it to a fixed cross-origin destination. The
transferred value changes the tool argument's payload, but not the
decision about which tool call or destination to use. The benchmark
evaluates cross-origin confidentiality violations, but not same-origin
flows, integrity attacks, or data injection. More broadly, WASP also
lacks data-dependent tasks, while WARP and DoomArena lack
confidential-data tasks. DoomArena covers both data-dependency classes,
but its BrowserGym evaluation treats an attack as successful when the
agent navigates to an attacker-controlled origin. It does not evaluate
disclosure of confidential data. Only MicroNet covers every
data-dependency class, origin boundary, attack goal, and attack
mechanism.

\end{document}